\documentclass[aps,prapplied,twocolumn,superscriptaddress,nofootinbib]{revtex4-2}
\usepackage{graphicx}
\usepackage{amssymb,amsfonts,amsmath}
\usepackage[colorlinks=true,citecolor=Cerulean,linkcolor=RubineRed,urlcolor=Cerulean]{hyperref}
\hypersetup{breaklinks=true}
\usepackage{graphicx}
\usepackage{color}
\usepackage[usenames,dvipsnames]{xcolor}
\usepackage{epstopdf}
\usepackage[normalem]{ulem}
\DeclareUnicodeCharacter{202F}{\,}
\usepackage{bm}
\usepackage{bbold}
\usepackage{float}
\usepackage{dsfont}
\usepackage{algorithm}
\usepackage{algpseudocode}

\renewcommand{\i}{{\rm i}}
\newcommand{\ket}[1]{|#1\rangle}
\newcommand{\bra}[1]{\langle #1|}

\renewcommand{\vec}[1]{{\bf #1}}

\begin{document}

\title{Quantum Secure Authentication with Color Centers in Diamond}

\author{Yannick Strocka}
\affiliation{Department of Physics, Humboldt-Universität zu Berlin, 12489 Berlin, Germany}

\author{Mohamed Belhassen}
\affiliation{Department of Physics, Humboldt-Universität zu Berlin, 12489 Berlin, Germany}

\author{Tim Schr{\"o}der}
\affiliation{Department of Physics, Humboldt-Universität zu Berlin, 12489 Berlin, Germany}
\affiliation{Ferdinand-Braun-Institut, Leibniz-Institut für Höchstfrequenztechnik, 12489 Berlin, Germany}

\author{Gregor Pieplow\textsuperscript{*}}
\affiliation{Department of Physics, Humboldt-Universität zu Berlin, 12489 Berlin, Germany}

\begin{abstract}

Quantum tokens offer a unique approach to secure authentication and payment in an increasingly networked, and eventually quantum-networked, world by enabling fully quantum-compatible authentication of users interacting with quantum computers at network nodes. A key component of token-based authentication is the reliable storage and verification of photonically transmitted token states, for example quantum states generated by a quantum physically unclonable function.
In this work, we introduce a quantum-compatible token architecture based on solid-state quantum memories using Group-IV color centers in diamond. Building on Wiesner’s quantum money framework, photonically generated token states are transmitted to a client and stored in a multi-qubit spin register, enabling delayed and non-destructive verification. We develop a realistic performance and security model that accounts for physical imperfections and optimal cloning attacks. Our results indicate that, with near-term improvements in device efficiency, high token acceptance rates are achievable for short-distance links.
These findings establish Group-IV color-center quantum memories as a promising platform for networked quantum authentication and token-based quantum applications beyond simple verification.
\end{abstract}

\maketitle

\section{Introduction}

\footnotetext{Corresponding author: gregor.pieplow@physik.hu-berlin.de}

Quantum networks across all scales have the potential to transform both computing \cite{Kimble2008,Kumar2025,Liu2025} and sensing \cite{Degen2017}. As quantum networks emerge \cite{Kimble2008}, the concept of quantum compatibility becomes increasingly relevant, with both new and existing routines \cite{bozzio_quantum_2025} benefiting from the ability to leverage a fully quantum network infrastructure. By quantum compatibility we refer to the ability of network routines to function entirely within a quantum ecosystem. In such a setting, a user could, for example, authenticate themselves \cite{Dutta2022} to initiate a blind quantum computing session \cite{Fitzsimons2017} and securely transmit instructions \cite{Gheorghiu2018} without relying on classical counterparts of these primitives. Adopting a quantum-compatible approach to such routines would not only exploit the intrinsic security guarantees of quantum cryptography but also enable the full use of the capabilities offered by quantum networks and their processing nodes.

Quantum authentication, in particular, offers a unique opportunity to leverage quantum effects for security purposes, for example in simple user authentication, and is especially important for protecting costly infrastructure from the theft of digital assets or identities. For example, quantum physically unclonable functions provide a promising approach to network authentication by generating transmissible quantum token states \cite{farre_secure_2025, ghosh_existential_2024}. More generally, the transmission and verification of such quantum tokens have attracted considerable attention \cite{pastawski_unforgeable_2012,georgiou_new_2015,bartkiewicz_experimental_2017,jirakova_experimentally_2019,Bilyk2023,mamann_quantum_2025}, particularly in the context of unforgeable quantum money \cite{Guan2018}, tokenized quantum signatures \cite{BenDavid2023}, and quantum tickets \cite{Pastawski2012}. One of the key challenges for quantum token–based authentication schemes, however, is the efficient and reliable storage of such tokens for practical applications.

In this work, we theoretically establish Group-IV vacancy color centers (G4V) as a robust platform for networked quantum-token-based authentication, thereby completing the blind quantum computing scenario outlined above. G4V centers, and in particular the silicon-vacancy (SiV) center, have already played a role in several key demonstrations, including a two-node quantum network \cite{Knaut2024,Nguyen2019}, entanglement-enhanced sensing \cite{stas_entanglement-assisted_2026}, and blind quantum computing \cite{wei_universal_2025}. Here we argue that G4V centers can also overcome efficiency bottlenecks \cite{strocka_memory_2025} that severely impact quantum token–based schemes. G4V centers possess long-lived spin qubits that form quantum registers capable of storing photonically transmitted token states through a highly efficient spin–photon interface \cite{OrphalKobin2024,strocka_memory_2025}.

We demonstrate, through detailed simulations, a protocol tailored to G4V centers coupled to an efficient spin–photon interface that incorporates realistic experimental constraints and enables full token storage and retrieval within Wiesner’s quantum money framework, thereby establishing the core storage capabilities required for network authentication based on quantum tokens. Wiesner’s quantum money is particularly attractive because it offers a simple yet robust scheme with well-established security bounds even under noisy conditions \cite{Wiesner1983}. Moreover, the token scheme demonstrates that the proposed memories can be reliably read out without destroying the stored quantum token.

\begin{figure*}[ht!]
    \centering
    \includegraphics[width=1.8\columnwidth]{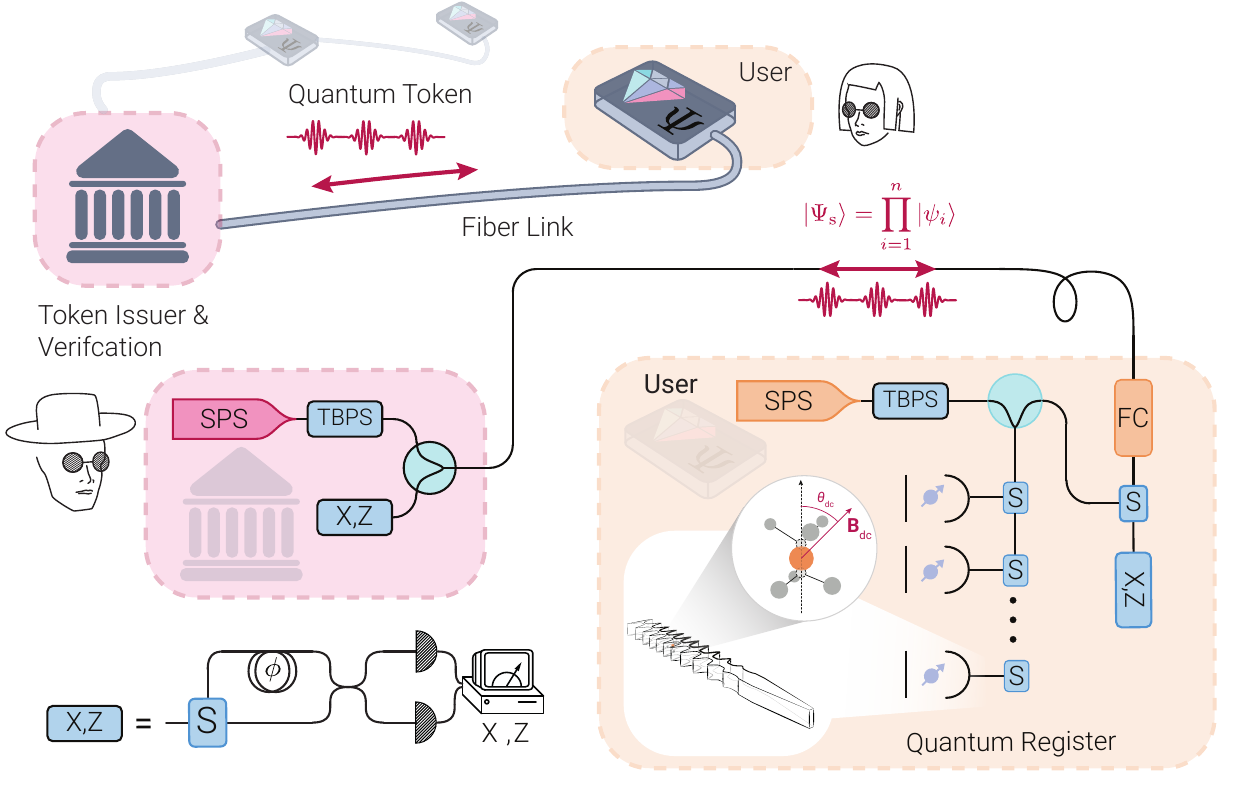}
    \caption{In our proposed quantum token scheme enabled by G4V, the process encompasses creation, storage, retrieval, and verification. Initially, the token issuer generates a quantum token $\ket{\psi_{\rm s}}$ along with a unique serial number $s$. The token is comprised of a sequence of photonic qubits $\ket{\psi_i}$, prepared using a single-photon source (SPS) and a time-bin qubit preparation stage (TBPS) \cite{Lee2018, Bouchard2022, Yu2025}. This configuration allows the issuer to encode each photonic qubit in the desired quantum state. The photons are generated in the infrared spectrum so that frequency conversion (FC) is required only at the user's end to enable interaction with the G4V. Upon transmission, the user routes and stores the token in a register composed of an array of high-efficiency sawfish \cite{bopp_sawfish_2024, pregnolato_fabrication_2024} spin-photon interfaces that host a G4V. Storage is achieved via either the electronic or nuclear spin. Fast switches (S) are deployed throughout the circuit to enable efficient measurement and routing of the token state. The $X$- and $Z$-basis measurements are performed using a fast switch integrated with an imbalanced Mach-Zehnder interferometer that features a controllable phase $\phi$ and two detectors (shown in the lower left corner). The storage procedure is completed once the photonic qubits are sequentially measured in the $X$-basis. For retrieval and verification, the spin state is read out by entangling it with a photon from an SPS, which is then directed to the verifier. A subsequent $Z$-basis measurement on the spin is performed, and verification succeeds if the retrieved token state sufficiently matches its original preparation (see Sec.~\ref{sec:token_scheme} for details). 
    }
    \label{fig:overview}
\end{figure*}

G4Vs are an attractive platform for quantum memories \cite{Bhaskar2020, Knaut2024, chen_scalable_2024, parker_diamond_2024}, as they preserve optical coherence in diamond nanostructures due to their reduced sensitivity to charge noise \cite{Bradac2019}. Our analysis accommodates heterogeneous single-photon sources and compensates for photon-to-photon frequency fluctuations (spectral diffusion \cite{ambrose_fluorescence_1991, acosta_dynamic_2012, orphal-kobin_optically_2023}) via optimized cavity parameters, enabling robust heterogeneous integration.

For token storage, we introduce high-fidelity optical Raman spin gates based on a pulse train of optical $\pi/8$ pulses and estimate a suitable magnetic field configuration given realistic cooperativity constraints. We further model the relevant physical limitations to derive practical design guidelines, focusing on the tin vacancy (SnV) \cite{Trusheim2020}, which offers longer coherence times \cite{harris_coherence_2023} than the silicon vacancy (SiV) \cite{hepp_electronic_2014} at comparable temperatures.

Finally, we derive the average token acceptance rate and evaluate achievable acceptance and redemption rates using realistic parameters while accounting for cloning attacks \cite{Molina2013}, finite temperatures, noise, and losses, thereby quantifying the technological readiness of the G4V platform for quantum-compatible applications.

\section{Token scheme} \label{sec:token_scheme}

Fig.~\ref{fig:overview} illustrates the high-level architecture of the token scheme. The token issuer (e.g., a bank) prepares a pair of a token state and serial number $\{\ket{\Psi_{\rm s}}, s\}$, where the token state is given by 
\begin{equation}
\ket{\Psi_{\rm s}} = \prod_{i=1}^n \ket{\psi_i}
\end{equation}
with the number of qubits $n$ and qubit $\ket{\psi_i}$, which is chosen uniformly at random from $\{\ket{0}, \ket{1}, \ket{+}, \ket{-}\}$ ($\ket{\pm}$ are the eigenstates of the Pauli $\sigma_x$ matrix). The serial number $s$ uniquely identifies the token without revealing the qubit states. We employ time-bin encoding so that 
\begin{equation}
\ket{\psi_i} = \alpha_i \ket{e_i} + \beta_i \ket{l_i}~,
\end{equation}
where $e_i$ and $l_i$ denote the early and late time bins of the $i$-th photon. The token state $\ket{\Psi_{\rm s}}$ is generated on demand using a single photon source (SPS) \cite{Reimer2019} in conjunction with fast switches and an imbalanced Mach-Zehnder interferometer (MZI) \cite{Shao2007, Yu2025}. The state is then transmitted via a fiber link to the user.

At the user side, the setup comprises an SPS, an imbalanced MZI, two single photon detectors, and a quantum memory register. In its simplest form, the register consists of $n$ sawfish cavities, each coupled to a G4V electron spin \cite{harris_coherence_2023}. The photons are sequentially routed to the corresponding memory spins. Alternatively, an electronic spin coupled to a nuclear spin may be used \cite{Harris2023}. Each incoming photon becomes entangled with the electronic spin via a spin-dependent reflection \cite{borregaard2020one}. The reflected photons are directed to a detector assembly featuring an imbalanced MZI and two single photon detectors. The sequence of detection events heralds the successful storage of the token state in the memory, thereby completing the token write procedure. This heralded mechanism not only ensures high fidelity in token storage but also facilitates the communication of lost qubits—an important feature for challenge-based verification schemes \cite{pastawski_unforgeable_2012, Bozzio2018}.

To demonstrate the full capabilities of the spin-diamond quantum memory register, the user can send the token back to the issuing entity for verification, as required in Wiesner's original scheme. This read-out process is achieved via a local SPS, an additional reflection-based spin-photon entanglement gate, and a subsequent spin projection measurement. When the issuer acts as the sole verifier, verification is performed by measuring the photons of the token state $\ket{\Psi_{\rm s}}$ in a basis determined by the serial number $s$ and comparing the outcomes with the tabulated expected results. Verification succeeds when a sufficient number of qubits match the expected values.

\section{Token creation, storage and retrieval}

The success of token verification depends on its creation, storage, retrieval, and verification. This section outlines the key assumptions and steps of the protocol.

\subsection{Token Creation}

We assume an on-demand SPS, such as a quantum dot \cite{Senellart2017}, color center \cite{Iwasaki2020}, or atom \cite{Basharov2010}. Either the SPS operates directly in the telecom C-band, which minimizes transmission losses and eliminates the need for frequency conversion, or a frequency converter is used to shift its central frequency to the telecom C-band. 
In this work, we assume that the emitter operates in the telecom C-band and that the photons can be converted to the optical range at the user's end. We require that the user can adjust the photons' central frequency $\omega_0$ to optimize the storage process.

The SPS photons have non-zero bandwidth (Fig.~\ref{fig:reflection_scheme}a), and a time-bin qubit preparation stage (TBPS) \cite{Lee2018, Bouchard2022} converts them at random into $\ket{e_i,l_i}$ or $\ket{\pm_i} = (\ket{e_i} \pm \ket{l_i})/\sqrt{2}$. The rapid phase modulation required to enable such swift changes is an active area of research \cite{Yu2025}. The corresponding phase settings are recorded and linked to a unique serial number $s$. 

We account for photon-to-photon fluctuations (spectral diffusion) but neglect multiphoton contributions, as weak coherent pulses and quantum dot double excitations have minimal impact. Notably, some closely related token schemes remain secure against multi-photon attacks \cite{pastawski_unforgeable_2012}.

\subsection{Token Storage and Retrieval}\label{sec:spe}

\begin{figure}[ht!]
    \centering
    \includegraphics[width=.74\linewidth]{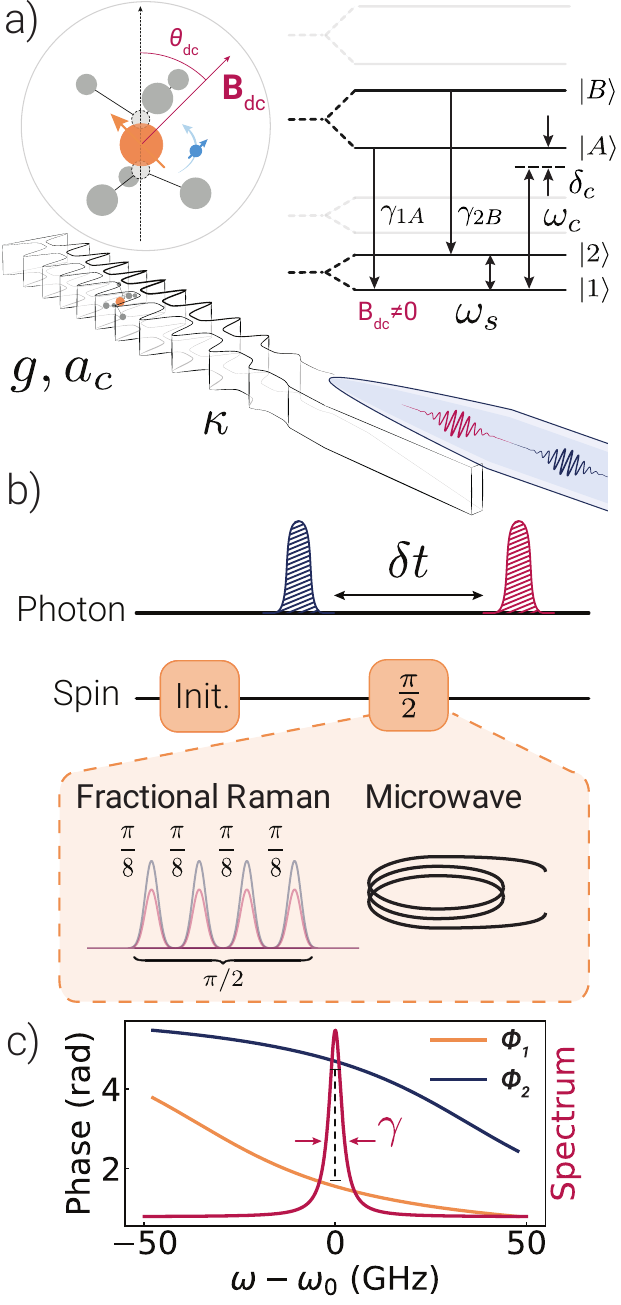}
    \caption{
    a) Spin-photon entanglement is mediated via the sawfish cavity-to-fiber interface \cite{bopp_sawfish_2024, pregnolato_fabrication_2024}, with the SnV center serving as a representative G4V system (electron spin in blue, nuclear spin in orange). For the reflection scheme, the SnV is modeled as a four-level system characterized by its spontaneous emission rates $\gamma_{1A},\gamma_{2B}$, spin splitting $\omega_s$, while interacting with a cavity mode $a_c$ at frequency $\omega_c$. b) Entanglement is generated using a reflection scheme in which the spin is initially prepared in $\ket{1}$, and the reflection of the incoming photon is spin-dependent. A $\pi/2$ rotation is then applied to the spin before the final reflection event. This rotation can be implemented either by a sequence of four $\pi/8$ optical Raman pulses or via microwave control. c) The figure displays the frequency-dependent phases $\phi_1(\omega)$ and $\phi_2(\omega)$ \cite{strocka_memory_2025}, along with the incident photon’s spectrum centered at $\omega_0$ and its bandwidth $\gamma$ at the frequency where $|\phi_1(\omega) - \phi_2(\omega)| \approx \pi$. Ideally, this phase difference remains constant across the photon’s spectrum; however, variations in $\phi_1(\omega)$ and $\phi_2(\omega)$ across the bandwidth introduce deviations. These deviations can be minimized by optimizing the parameters $\omega_0$, $g$, $\omega_c$, and $\kappa$ to enhance the fidelity of the spin-photon entanglement. An additional control sequence is required to transfer the spin state onto the nuclear spin for readout and storage.
    }
    \label{fig:reflection_scheme}
\end{figure}

Token storage relies on a phase gate between individual spins and photons using a spin-dependent reflection scheme at the sawfish cavity interface \cite{borregaard2020one,strocka_memory_2025,strocka_repeater_2025}, followed by a projective measurement that heralds the stored state (Fig.~\ref{fig:reflection_scheme}). The detailed procedure is provided in App.~\ref{app:singlefre}.

We assume a half-open cavity with negligible internal losses, which we incorporate into the cavity-to-fiber coupling efficiency \cite{bopp_sawfish_2024}. 

As shown in Fig.~\ref{fig:reflection_scheme}, after initializing the spin in $\ket{1}$, an incident photon $\alpha\ket{e} + \beta \ket{l}$ interacts with the spin via state-dependent reflection, separated by a $\pi/2$ rotation between time bins. Ideally, the reflection phase is perfectly correlated with the spin state, leading to transformations such as $\ket{e,l}\ket{1} \rightarrow -\ket{e,l}\ket{1}$ and $\ket{e,l}\ket{2} \rightarrow \ket{e,l}\ket{2}$. This correlation entangles the spin and photon. A measurement of the photonic qubit in the $X$-basis then heralds the storage of the quantum state in the memory as $\alpha \ket{1} \pm \beta \ket{2}$, up to a known spin-state rotation. In the full protocol, this rotation is unnecessary if measurement outcomes are communicated to the token verifier (App.~\ref{app:singlefre}).

The spin state can also be transferred to a nuclear spin \cite{hanson2024}, enabling longer coherence times crucial for long-distance transmissions \cite{Wang2025} and improved token validation for extended storage durations.

State retrieval follows the same procedure: an SPS, together with a time-bin qubit preparation stage (TBPS), generates a photon in the state $(\ket{e} + \ket{l})/\sqrt{2}$. This photon is reflected off the spin with another $\pi/2$ rotation between time bins before being sent back to the issuer. A final spin measurement in the $Z$-basis heralds the restored photonic quantum state. Storage and retrieval measurement results are then transmitted to the verifier along with the qubit.

In an idealized scenario, incoming photons have a spectral width much narrower than the cavity response, ensuring a phase contrast of
\begin{align}\label{eq:contrast}
\Delta\phi(\omega) := \phi_{1}(\omega)-\phi_{2}(\omega) = \pi
\end{align}
across the photon spectrum. However, as shown in Fig.~\ref{fig:reflection_scheme}a, variations in $\Delta\phi(\omega)$ due to the photon bandwidth may reduce controlled phase (CP) gate fidelities.

To mitigate these errors, the cavity response can be optimized by tuning the cavity mode central frequency $\omega_c$ and the cavity loss rate $\kappa$ \cite{Omlor2025}. This optimization is particularly important for high-rate SPSs, such as quantum dots \cite{Senellart2017}, which emit broader-bandwidth photons than the cavity emitter’s natural linewidth $\gamma_{1A}$. Furthermore, as will be discussed in Sec.~\ref{sec:qual}, careful cavity design allows tolerance to a certain level of spectral diffusion in the SPS.

The following sections detail the optimization of the cavity design for non-zero bandwidth photons, fabrication uncertainties, and coherent spin control. Finally, we characterize the overall performance of the token scheme using the optimized control and cavity parameters.

\subsection{Coherent Control}\label{sec:raman}
Both reading and writing the token state make use of a $\pi/2$ rotation around the $y$-axis on the Bloch sphere. This rotation can be implemented via microwave control \cite{rosenthal_microwave_2023, karapatzakis_microwave_2024, pieplow_efficient_2024} or through all-optical Raman control \cite{Becker2018, debroux_quantum_2021, Takou_2021, pieplow_deterministic_2023}. Each approach has distinct trade-offs. Microwave control has demonstrated high-fidelity rotations; however, it typically necessitates either a highly strained environment \cite{rosenthal_microwave_2023, karapatzakis_microwave_2024} or a special magnetic field configuration \cite{pieplow_efficient_2024}. 
Optical Raman control poses a greater challenge for achieving high-fidelity gates under low-strain conditions \cite{pieplow_deterministic_2023}, yet it can bridge ground state splittings $\omega_s$ that exceed the frequency range of commercially available microwave equipment. Moreover, because Raman control can handle larger ground state splittings, its quantum speed limit is substantially lower than that of microwave control—thereby enhancing transmission rates, as discussed in Sec.~\ref{sec:qual}. To highlight their complementary strengths, we incorporate both control schemes in our analysis.

{\emph{Raman control:}} A comprehensive explanation of an all‐optical Raman control scheme is provided in \cite{pieplow_deterministic_2023}. In \cite{pieplow_deterministic_2023}, we show how two laser pulses with central frequencies $\omega_1$ and $\omega_2$, both detuned from the excited states, can be employed to generate high-fidelity Raman spin gates.
Achieving such high fidelity requires precise tuning of several parameters: the magnetic field orientation (which lifts the spin degeneracy), the laser field strengths, the pulse duration, the relative phase, the polarization, and the detuning. As noted in \cite{pieplow_deterministic_2023}, the primary challenge in implementing these optical spin gates is minimizing the transient population in the excited states.

In this work we extend the findings of \cite{pieplow_deterministic_2023}. Instead of implementing a $\pi/2$ gate with a single pulse, we optimize control using a pulse train in which each pulse produces a fractional rotation. As illustrated in Fig.~\ref{fig:reflection_scheme}b, these fractional rotations minimize transient excited-state populations compared to a single, $\pi/2$ rotation. In our approach, a sequence of four $\pi/8$ rotations around the $y$-axis of the Bloch sphere collectively yields the desired $\pi/2$ gate. A detailed explanation of the optimization process and relevant system parameters is provided in App.~\ref{app:spingates}, and the optimization results are summarized in Tab.~\ref{tab:table_ext}.

We report a $\pi/2$ gate fidelity of $F^{\rm R}_{\pi/2} = 0.9978$ for a pulse train with a total duration of $T_g = 1.9$ ns and an interpulse-delay $\Delta\tau=T_g/4=475.1$ ps, using an SnV center in a low-strain environment at a magnetic field of $B_{\rm dc} = 3.0$ T and temperature $T = 0.1$ K. A magnetic field strength of $3.0$ T was chosen because it minimizes fidelity degradation in the optical spin gates \cite{pieplow_deterministic_2023}. While the $\pi/2$ gate is robust against variations in system parameters including the interpulse-delay, pulse amplitudes and polarization, precise control of the magnetic field orientation and pulse phase difference is essential to maintain a fidelity above $0.9950$. A detailed discussion is provided in App.~\ref{app:robustness}.

The reported gate fidelity accounts for both photonic and phononic relaxation, as well as phononic dephasing. The phononic contribution to the overall decoherence is calculated following the approach in \cite{harris_coherence_2023}, which explicitly includes the dependence on magnetic field orientation and strength.

\emph{Microwave control:} In \cite{pieplow_efficient_2024}, it was predicted that an optimal magnetic field configuration exists for efficiently controlling the two lowest spin states of a G4V defect, regardless of strain. For our token scheme, we adopt this configuration, where the static magnetic field is oriented orthogonal to the defect's polarization axis and the microwave field is polarized along its symmetry axis. At $0.1$ K, $B_{\rm dc}=139$ mT and $B_{\rm ac}=1.0$ mT (details are provided in Sec. \ref{sec:robust}) we obtain a microwave gate fidelity of $F^{\rm MW}_{\pi/2} = 0.9998$ with a gate duration of $T_g = 15.56$ ns. Our analysis also includes the effects of phononic dephasing, which depend on the field direction.

\subsection{Cavity Requirements}\label{sec:robust}

We now describe how to calculate the cavity parameters $\,\omega_c$ and $\kappa$, as well as the incident photon's central frequency $\omega_0$, at a given magnetic field strength $B_{\rm dc}$ and orientation $\theta_{\rm dc}$, to maximize the CP gate fidelity. We assume that $\omega_0$ can be chosen during the frequency conversion step.

For both Raman and microwave control, the field strength and orientation are determined by the requirements of each control scheme (e.g., $\theta_{\rm dc}$ in Tab.~\ref{tab:table_ext}). The magnetic field sets the relaxation and dephasing rates (see App.~\ref{app:phonons}). Both the field strength and orientation also determine the splitting between the ground and excited states.

We first obtain an initial estimate for the cavity parameters and the incident photon's frequency $\omega_0$ 
by optimizing the spin–photon entanglement fidelity under the assumption of negligible crosstalk, following the method described in \cite{strocka_memory_2025}. In this step, we model the incident photon with a Lorentzian intensity profile. As detailed in \cite{strocka_repeater_2025}, this initial estimate then serves as the starting point for the full optimization of the coupled Heisenberg–Langevin equations derived in \cite{strocka_memory_2025}, in which we explicitly account for crosstalk involving undesired transitions.

Maximizing the fidelity with respect to $(\kappa,\omega_c,\omega_0)$ yields a set of optimal cavity design parameters ensuring that photons with a non-zero bandwidth $\gamma$ experience the desired phase shift over the relevant frequency range.

Here, we characterize the light–matter interaction strength through the cooperativity for the transition $\ket{1}\leftrightarrow\ket{A}$, i.e. \(C = \vert g_{1A}\vert^2/(2\kappa\gamma_{1A})\), which quantifies the ratio of coherent coupling \(g_{1A}\) to the cavity decay rate \(\kappa\) (HWHM) and the emitter’s linewidth \(\gamma_{1A}\) (FWHM) \cite{strocka_memory_2025}.
Notably, a cooperativity of $C = 25$ has been demonstrated for group-IV color centers in diamond, specifically for the SiV center \cite{Bhaskar2020}. For the SnV center, however, the experimentally reported cooperativities are still lower \cite{Herrmann2024}. Since both SiV and SnV belong to group-IV color centers, they are expected to reach comparable performance in the near future \cite{nina2025}. Therefore, we adopt $C = 25$ as a realistic, forward-looking parameter. While this choice yields an upper bound on the achievable fidelity \cite{strocka_repeater_2025}, the present quantum-token scheme remains compatible with narrow-band photon pulses and is therefore not fundamentally limited by this assumption.

For the reported optimized magnetic field orientations for optical control we optimize the fidelity of the phase gate assuming an incoming photon with the bandwidth $\gamma=600$ MHz. For the magnetic field configuration $B_{\rm dc}=3.0$ T and $\theta_{\rm dc}=43.11\deg$ which produces the highest spin gate fidelity for optical control we find $1-F_{\rm CP}=0.0298$ and $\eta_{\rm CP}=0.9895$. Assuming the same optimized parameters the infidelity is a linear function of the bandwidth while the efficiency stays the same (see App. \ref{app:cavity} for details). 

For microwave control we are not restricted to a specific magnetic field strength or orientation \cite{pieplow_efficient_2024}. However, the cooperativity limit \(C = 25\) imposes a natural optimum on the DC magnetic-field strength for any fixed field orientation and incident-photon bandwidth. 

Increasing the magnetic field enhances the optical splitting \cite{strocka_repeater_2025,omlor_entanglement_2024}, but this does not indefinitely improve the fidelity. Beyond a certain point, the cooperativity constraint prevents further gains, thereby setting an upper bound on the achievable performance. Specifically, there exists an optimal optical splitting because excessively large splittings separate the spin-dependent phase responses such that they no longer produce the wide frequency range in which the phase condition shown in Eq. \eqref{eq:contrast} is ideally met (see App.~\ref{app:cavity} for details).  

To illustrate that behavior we numerically optimize the fidelity \cite{strocka_memory_2025}
\begin{align}
    F_{\rm CP}=\langle\psi\vert\rho\vert\psi\rangle
\end{align}
with $\ket{\psi}=1/\sqrt{2}(\ket{1}+\ket{2})$ and the spin state $\rho$ after measurement over a range of magnetic-field strengths. At the resulting optimal field strength, we then perform a local optimization in the presence of crosstalk over the parameters \((B_{\rm dc}, \delta_c, \delta_0)\) with $\delta_0:=\omega_{1A}-\omega_0$ and $\delta_c=\omega_{1A}-\omega_c$. This yields \(1 - F_{\rm CP} = 0.0204\) and \(\eta_{\rm CP} = 0.9718\) at the optimized field strength \(B_{\rm dc} = 139\ \mathrm{mT}\) and orientation \(\theta_{\rm dc} = \pi/2\). The optimized infidelity and the corresponding efficiency remain robust against cavity perturbations (see App.~\ref{app:cavity} for details).

\section{Token performance}\label{sec:qual} 

We now evaluate the anticipated performance of the token scheme under the optimal token storage and retrieval configuration, integrating these results with a security analysis. A system-aware security analysis is essential for fully assessing the performance of our token-memory scheme.

Analogous to the secret bit rate in communication protocols, we define an expected token acceptance rate for a series of tokens that are issued, stored, retrieved, and returned to the verifier. This acceptance rate depends on a security parameter that sets the minimum token length required to ensure that, even in the presence of losses and noise, the probability of a successful adversarial attack is negligible \cite{Molina2013}. In our case, the rate critically depends on transmission, conversion, and interface losses; token storage time; the type of storage spins (electronic or nuclear); the control scheme; source bandwidth; and imperfections such as spectral diffusion of the SPS that emits the token.

We consider an optimal cloning attack \cite{Molina2013} on Wiesner's scheme as the worst-case scenario. Although perfect cloning is forbidden by the no-cloning theorem, imperfect copies of a single-qubit state can be made \cite{bartkiewicz_experimental_2017} and used to forge a token.

Let $p_{\rm af}$ denote the probability of successfully accepting a forged token. For an optimal cloning attack without losses, \cite{Molina2013} gives 
\begin{equation}
    p_{\rm af} = \alpha^n,
\end{equation}
where $n$ is the number of qubits in the token state and $\alpha = 3/4$ is the single-qubit cloning probability for Wiesner's original scheme.

In the presence of losses, the probability of accepting a forged token becomes \cite{Molina2013}
\begin{equation}
    p_{\rm af}(n,t)=\sum_{k=t}^{n} \begin{pmatrix}
        n\\ k
    \end{pmatrix}\alpha^k\left(1-\alpha\right)^{n-k},
\end{equation}
where $t$ is the minimal number of successful measurements by the verifier out of a total of $n$ qubits in the token. 

We define a security threshold $p_{\rm th}$ by requiring that
\begin{equation}
    p_{\rm af}(t,n) < p_{\rm th}.
\end{equation}
This condition identifies the smallest token size $n$ (with at least $t<n$ successful measurements) for which the probability of accepting a fake token is below $p_{\rm th}$. In our simulations, we consider thresholds of $p_{\rm th}=10^{-4},\;10^{-5},\;10^{-6}$, which can be adjusted based on the security requirements and the volume of tokens used. These thresholds will also help to illustrate how the token acceptance rate depends on the desired security level.

Tab.~\ref{tab:thres} lists the corresponding values for $t$ and $n$, confirming that higher security demands require larger token sizes. Notably, token sizes as small as 59 qubits can reduce the probability of a forgery to one in a million. We also note that other token schemes with more relaxed bounds for $\alpha$ \cite{Molina2013} could further reduce the required token size.
\begin{table}[htb!]
\begin{center}
\caption{Probability thresholds with corresponding token sizes}
    \centering
    \begin{tabular}{ccc}
        \toprule
         $p_{\rm th}$ & $n$ & $t$\\
         \hline
         $10^{-4}$ & $42$ & $41$\\
         $10^{-5}$ & $51$ & $50$\\
         $10^{-6}$ & $59$ & $58$\\
    \toprule
    \end{tabular}
    \label{tab:thres}
\end{center}
\end{table}
So far we have only considered the probability of accepting a forged token without accounting for the chance of a true positive verification. The probability of successfully accepting $k$ out of $n$ qubits in a token state is given by
\begin{equation}
    p_a(k,\langle F\rangle)=\begin{pmatrix}
        n\\ k 
    \end{pmatrix}\langle F\rangle^k (1-\langle F\rangle)^{n-k}~.
\end{equation}
The average fidelity is defined as $\langle F\rangle=\frac{1}{4}(F_++F_-+F_e+F_l)$, where $F_x$ for $x=\pm,e,l$ accounts for both the storage and retrieval processes which include the effects of rotation gates and the non-zero bandwidth of the incident photons as well as the storage duration which may introduce additional dephasing. We elaborate on the evaluation of $F_x$ in App. \ref{app:performance}.

The error channels such as radiative decay and phonon induced dephasing in the color center as well as the bandwidth of the incident photon from the single photon sources are provided in \cite{strocka_memory_2025}.

With the true positive token verification probability, we define the average acceptance rate as
\begin{align}\label{eq:acceptance_rate}
    \gamma_a= \gamma_{\rm tok} \sum_{k=t}^{n} p_a(k,\langle F\rangle) p_{\rm loss}(n, k)
\end{align}
where $\gamma_{\rm tok}$ is the token processing rate and
\begin{equation}
   p_{\rm loss}(n, k) = \begin{pmatrix}
        n\\ k
    \end{pmatrix} p_1^k (1-p_1)^{n-k}
\end{equation}
is the probability of losing $k$ out of $n$ photons during the transmission of the token. 
The single photon loss probability is $p_1 = \langle\eta\rangle\eta_c e^{-L/L_{\rm att}}$, where $\langle\eta\rangle=\frac{1}{4}(\eta_++\eta_-+\eta_e+\eta_l)$ is the average efficiency of the storage and retrieval process (see App. \ref{app:performance} for details), $\eta_c$ accounts for all the relevant interface and device efficiencies, $L$ is the transmission fiber length and $L_{\rm att}$ its attenuation length. The combined interface and device efficiency is $\eta_c=\eta_{\rm cf}^2\eta_{\rm fc}^2\eta_{d}^2$.

The processing rate is defined as $\gamma_{\rm tok} = 1 / T_n$, assuming that the reading and writing processes require the same amount of time. The total processing time is given by
\begin{equation}
    T_n = 2 n(T_{\rm tb} + T_g + T_m) + 2 T_c + T_s,
\end{equation}
where $T_{\rm tb}$ is the time allocated for a time-bin qubit, $T_g$ is the control gate duration, $T_m$ is the measurement time, $T_s$ is the storage time, and $T_c$ is the transmission time. We set $T_{\rm tb} = 20 T_{\rm lt}$, where $T_{\rm lt}$ is the lifetime of the single photon source, and choose $T_g = 40\sigma$ with 
$
\sigma = \tau_{\pi/8}/2\sqrt{2\ln(2)}
$,
where $\tau_{\pi/8}$ is the full width at half maximum of the optimized $\pi/8$ pulse for $B_{\rm dc}=3.0$ T (see Tab.~\ref{tab:table_ext}). If microwave control is chosen the gate duration $T_g$ is a quarter of a full Rabi oscillation \cite{pieplow_efficient_2024}. We take $T_m = 100$ ps \cite{cherednichenko_low_2021}, corresponding to the dead time of the photon detectors \cite{Grotowski2025}. The communication time is given by $T_c = L / c$, where $L$ is the communication distance and $c=2\cdot 10^8$ m/s is the speed of light in the fiber.

In Eq.~\eqref{eq:acceptance_rate}, the security threshold $p_{\rm th}$ determines the values of $n$ and $t$. Importantly, the fidelity sets the frequency of true positive events; the higher the fidelity, the higher the average acceptance rate $\gamma_a$.

\subsection{Results}
\begin{figure*}[htb!]
    \centering
    \includegraphics[width=\linewidth]{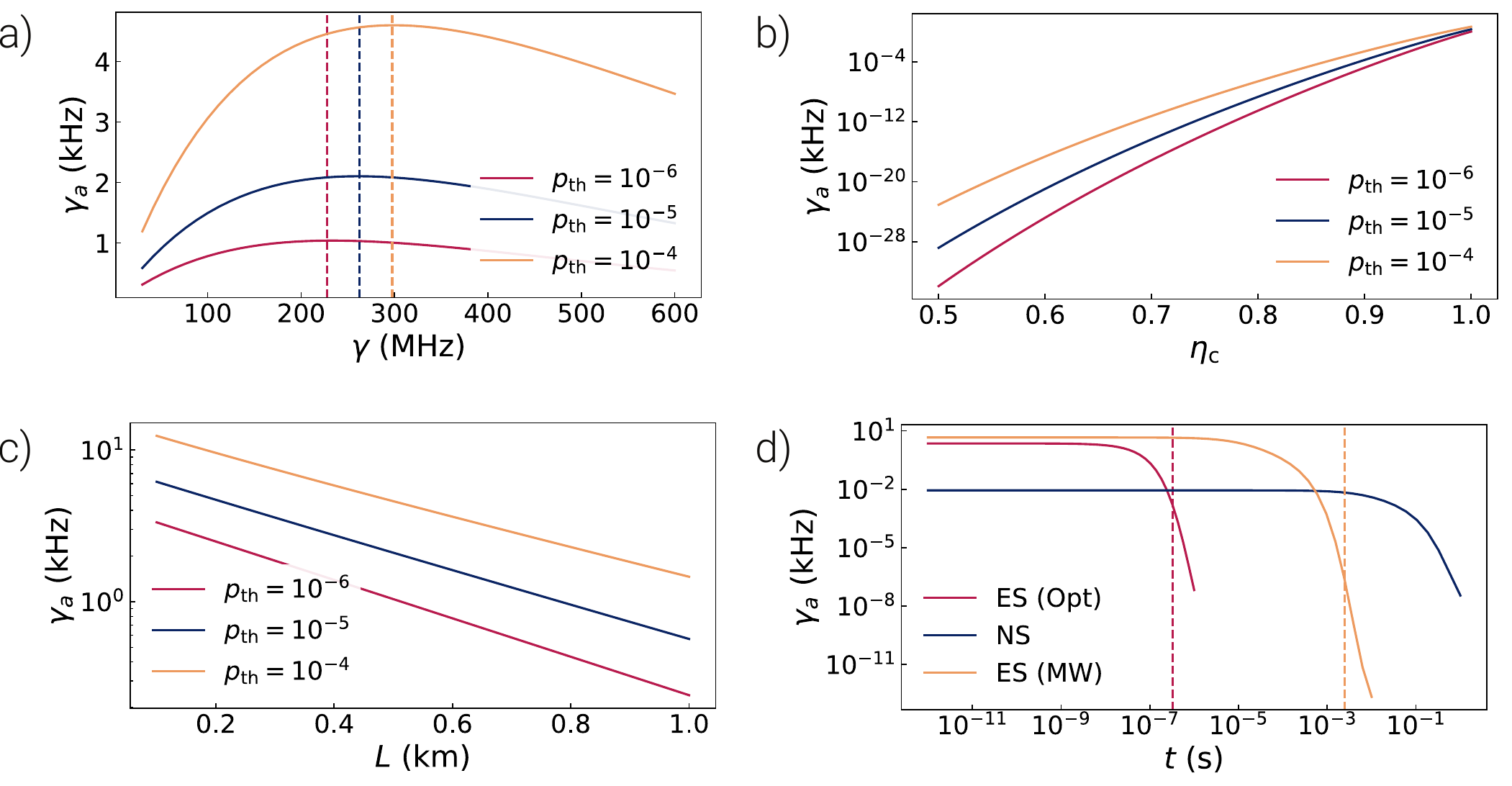}
    \caption{Token acceptance rates. Unless stated otherwise, graphs a)–d) assume a fiber length of $L=0.5$ km, attenuation length $L_{\rm att}=20$ km, operating temperature $T=0.1$ K, error probability threshold $p_{\rm th}=10^{-4}$, and cavity coupling efficiency $\eta_c=1.0$. Storage time is zero except in d). A $\pi/2$ gate is implemented via microwave control. a) The acceptance rate $\gamma_a$ is shown as a function of the incoming photon’s bandwidth $\gamma$ for $p_{\rm th}=10^{-4},10^{-5},10^{-6}$. Dashed lines indicate the bandwidth yielding the highest acceptance; for $p_{\rm th}=10^{-4}$, $\gamma_{a,\rm max}=4.60$ kHz is achieved at $\gamma=298$ MHz. b) $\gamma_a$ versus $\eta_c$ is plotted at the optimal bandwidth [dashed lines in a)], revealing a rapid decay in rate for $\eta_c < 0.9$. c) $\gamma_a$ is plotted as a function of fiber length $L$ at the optimal $\gamma$, indicating that $L$ is not the limiting factor within the considered range. d) $\gamma_a$ is illustrated as a function of memory time $t$ for the electron spin (ES) controlled optically (Opt) or via microwaves (MW), and for the nuclear spin (NS) at $T=0.1$ K. The initial rates are $\gamma_a(0)=2.27$ kHz for ES (Opt), $\gamma_a(0)=4.60$ kHz for ES (MW), and $\gamma_a(0)=8.80$ Hz for NS. The dashed lines denote the time $t$ for which $\langle F(t)\rangle$ crosses the threshold $3/4$.}
    \label{fig:acceptance_rate}
\end{figure*}

The token acceptance rate $\gamma_a$ is the key performance indicator of the scheme, allowing us to assess the diamond spin-memory register while accounting for the main error sources and an advanced attack scenario.

For our simulations, we assume a magnetic field strength of $B_{\rm dc}=0.139$ T for microwave control with the field orientation $\theta_{\rm dc}=\pi/2$ and a driving field $B_{\rm ac}=10^{-3}$ T with $\theta_{\rm ac}=0$, and make use of the optimized parameters from Tab. \ref{tab:CP_data} in App. \ref{app:cavity} for $\gamma=600$ MHz of a SPS. 
We assume a temperature of $T=0.1$ K and a total communication distance of $L = 0.5$ km, extendable with quantum repeaters \cite{Wo2023}. Apart from that, we assume that the fidelity of photon generation for token storage and retrieval are both $F_{\rm ph}=0.99$. The parameters are all compatible with recent experimental efforts for both G4Vs as well as SPS \cite{rosenthal_microwave_2023,Kuhlmann2015}. 

In Fig. \ref{fig:acceptance_rate} we study $\gamma_a$ and its dependence on critical system parameters. Fig. \ref{fig:acceptance_rate}a shows the dependence on the photon's bandwidth (assuming $T_s = 0$ and $\eta_c = 1$). Here, an increasing photon bandwidth $\gamma=1/(T_{\rm lt})$ boosts the token generation rate $1/T_n$, but reduces the token fidelity due to imperfect bandwidth matching. The optimal bandwidth is at $298$ MHz (corresponding to a SPS lifetime $T_{\rm lt}= 3.36$ ns) and yields $\gamma_a = 4.60$ kHz for $p_{\rm th}=10^{-4}$. Such lifetimes require only moderate Purcell enhancement for quantum dots and are within reach for G4V \cite{bopp_sawfish_2024}.

Fig. \ref{fig:acceptance_rate}b presents the token acceptance rate $\gamma_a$ as a function of the total device efficiency $\eta_c$ at $\gamma_{\rm opt}$ for each $p_{\rm th}$. With current state-of-the-art values $\eta_c \approx 0.4915$ (from $\eta_{\rm cf}=0.98$, $\eta_{\rm fc}=0.73$, and $\eta_d=0.98$), $\gamma_a$ is essentially zero. However, if near-future improvements bring $\eta_c > 0.9$, the acceptance rate $\gamma_a$ will start exceeding the Hz range.

Fig. \ref{fig:acceptance_rate}c shows the rate $\gamma_a$ as a function of the fiber length $L$ for $T_s = 0$ and $\eta_c = 1$ at $\gamma_{\rm opt}$, with $\gamma_a > 1.46$ kHz for $L<1$ km for all $p_{\rm th}$.

Finally, Fig. \ref{fig:acceptance_rate}d depicts $\gamma_a$ as a function of the storage time $t$ for three scenarios: storage using the electron spin (ES) with either optical (Opt) or microwave (MW) control gates, and storage using nuclear spins (NS) (with microwave control) at $T=0.1$ K. For optical control we assume $B_{\rm dc}=3.0$ T, $\theta_{\rm dc}=43.11°$ and $T=0.1$ K as shown in Tab. \ref{tab:table_ext} of App. \ref{app:spingates}. Spin dephasing due to phonons is given in App.~\ref{app:spin_dec}, and the electronic spin decay and excitation rates in App.~\ref{app:phonons}. For the nuclear spin, we assume  microwave control and a swap fidelity using radio waves of $0.9993(5)$ with a gate duration $T_g=0.1$ ms \cite{bartling} and a nuclear spin dephasing time of $T_2=1$ s \cite{grimm_coherent_2025} (see App.~\ref{app:spin_dec}).

The electron spin controlled optically has an initial rate of approximately $2.27$ kHz, dropping to nearly zero after about $1$ $\mu$s, while the microwave-controlled electron spin starts at around $4.60$ kHz and decreases significantly after about $1$ ms. The longer coherence time of the microwave-driven electron spin is due to a field-dependent coherence time, which is reduced for the optimal $\theta_{\rm dc}$ for optical control. The nuclear spin memory exhibits the longest coherence time but the lowest initial acceptance rate ($8.80$ Hz, exponentially decreasing after more than $10$ ms).

For optical control we find an average fidelity and efficiency for storage and retrieval $(\langle F\rangle,\langle\eta\rangle)= (0.9205,0.9703)$ at the optimal $\gamma = 216$ MHz, for microwave control the optimal $\gamma = 298$ MHz results in  $(\langle F\rangle,\langle\eta\rangle) = (0.9652,0.9498)$,  and for nuclear spin storage $\gamma = 20$ MHz implies $(\langle F\rangle,\langle\eta\rangle) = (0.9833,0.9530)$.

In summary, for short storage times the electronic spin is preferable, with microwave control performing best for short durations, while for longer storage times the nuclear spin is the better choice. Optical control has the advantage of the highest token processing rate $\gamma_{\rm tok}$ due to its fast spin gate. Note that these results assume a worst-case attack scenario; a weaker adversary or a scheme requiring fewer photons would yield a higher $\gamma_a$. A challenge–response scheme \cite{pastawski_unforgeable_2012} would already guarantee an improvement, because the token only has to be transmitted once.
\begin{figure}[tb]
    \centering
    \includegraphics[width=\linewidth]{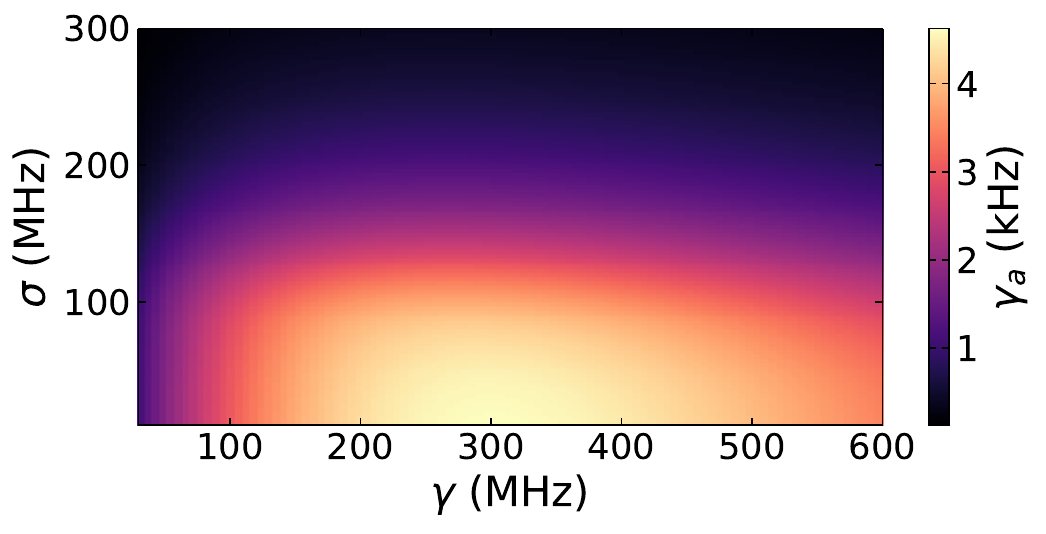}
    \caption{Spectral diffusion's standard deviation $\sigma$ impact on $\gamma_a$ for different $\gamma$ of the incoming photons is quantified. We assume $L=0.5$ km, $L_{\rm att}=20$ km, $T=0.1$ K, $p_{\rm th}=10^{-4}$, $\eta_c=1$, microwave control to achieve a spin $\pi/2$ rotation (see details in Tab. \ref{tab:table_ext} from App. \ref{app:spingates}), the electron spin as the quantum memory, $T_s=0$ and a gaussian shape of the spectral diffusion with standard deviation $\sigma$.}
    \label{fig:spec_diff}
\end{figure}

Finally, we examine the impact of spectral diffusion of the SPS. While spectral diffusion can limit entanglement distribution \cite{barrett_efficient_2005}, since remote photons must interfere, it is less problematic here because the incident photons only need to interfere with themselves for storage.
In Fig.~\ref{fig:spec_diff}, we show the dependence of $\gamma_a$ on the bandwidth of incoming photons and a random distribution of their central frequencies. This distribution is modeled as a normal distribution centered at $\omega_0$ with standard deviation $\sigma$, in agreement with experimental observations \cite{OrphalKobin2024} and previous theoretical studies \cite{kambs_limitations_2018}. For evaluating $\gamma_a$ in Fig.~\ref{fig:spec_diff} we explicitly calculate the density matrix of the stored state in the presence of spectral diffusion (see App. \ref{app:diff}). 

The results indicate that $\gamma_a$ only weakly changes for $\sigma<100$ MHz. This suggests that emitters operating in this regime, including the SnV center \cite{trusheim_transform-limited_2020}, are compatible with our approach and that the scheme exhibits substantial robustness against spectral diffusion, an important feature of our present quantum token proposal.

\section{Summary and Outlook}\label{sec:sum}

Our practical quantum token scheme, which leverages G4V centers for secure token storage and retrieval, shows that quantum-protected and functional scenarios are feasible in the near future.

By integrating robust spin‑photon interfaces, high‑fidelity $\pi/2$ gates (realized via optimized pulse trains), and carefully engineered cavity parameters, we show that, even with practical constraints such as non-zero photon bandwidth and fabrication uncertainties, it is possible to achieve exceptional token storage and retrieval fidelities. 

Our analysis indicates that with state‑of‑the‑art device parameters and near‑term improvements in conversion and coupling efficiencies, token acceptance rates can reach the kHz regime under optimal conditions. The rates shown in Fig.~\ref{fig:acceptance_rate} are mainly limited by the communication distance. We estimate a processing duration of approximately $150$ ns per qubit, excluding transmission time. By parallelizing the processing of individual photonic qubits through frequency and spatial multiplexing \cite{Komza2025}, and by using multiple detectors and sources on both the issuer and user sides, the acceptance rates can be drastically increased—potentially reaching the MHz regime—making the scheme highly promising for secure quantum applications.

Our performance analysis is based on detailed modeling of the G4V (specifically the SnV), including all relevant interactions and major error sources such as undesired spontaneous relaxation and phononic dephasing. We perform complex parameter optimizations that yield robust optimal parameters for both the cavity and the control operations. Notably, the token acceptance rate is remarkably robust against spectral diffusion, since our reflection scheme accommodates wide-bandwidth photons—an advantage for both token and repeater applications.

Regarding control, we compare optical and microwave strategies and emphasize our use of fractional Raman gates to achieve the desired rotations. Although fractional gates maximize fidelity at the expense of speed, their detailed analysis is an interesting direction for future research.

Overall, these developments not only pave the way for the experimental realization of secure quantum token systems but also lay a solid foundation for integrating such protocols into larger-scale quantum networks. Future work may consider the design of a quantum register for large GHZ states \cite{Cao2024} for enhancing security towards cloning attacks \cite{Molina2013}. Given the proximity of the architecture to quantum repeater proposals it is straightforward to include advanced error correction techniques, that protect against photon losses \cite{borregaard2020one} as well as logical errors \cite{Wo2023}. 

Finally, we want to emphasize that the quantum memories are not limited to state storage; when integrated with a ${}^{13}$C spin register \cite{Bradley2019}, they can also function as small quantum processors. This integration enables in-situ quantum information processing and supports advanced protocols such as quantum identification schemes based on physically unclonable functions. In this way, our proposed register not only provides robust memory capabilities but also serves as a building block for fundamentally new approaches to quantum security. Altogether, the unique properties and versatility of our design make it an exciting platform for future quantum security applications.

\section*{Acknowledgements}

Funding for this prject was provided by the German Federal Ministry of Education and Research (BMBF, project QPIS, No. 16KISQ032K, QPIS.2, No. KIS6GCQ020; ERC StG Grant QUREP of the EC, No. 851810, ERC Con. Grant HyperGraph, No. 101171255).

\section*{Author Contributions}

Y.S. and G.P. conceptualized the research, with Y.S. conducting the simulations and data analysis. M.B. provided the code specific to microwave spin control. G.P. and T.S. developed the core idea and provided overall project supervision. All authors contributed to writing and refining the manuscript.
\bibliographystyle{aipnum4-1}
\bibliography{token_npj.bib}

\cleardoublepage

\begin{appendix}

\section{READING PROCESS}\label{app:singlefre}
\indent The writing process consists of the first reflection, a $\pi/2$ rotation and a second reflection \cite{borregaard2020one}. The reading process works the same way. Let's assume single frequency photons, i.e. $\ket{e}=\ket{l}=e^{{\rm i}\omega_0 t}$, a half-open cavity which yields a unity amplitude of the reflectivity and $\phi_\downarrow(\omega_0)=\pi$, $\phi_\uparrow(\omega_0)=0$ for simplicity. The starting point is the saved spin state at the quantum memory
\begin{align}
    \langle +\vert\psi\rangle=\frac{1}{2}(-\alpha-\beta)\ket{{1}}+\frac{1}{2}(-\alpha+\beta)\ket{{2}},\\
    \langle -\vert\psi\rangle=\frac{1}{2}(-\alpha+\beta)\ket{{1}}-\frac{1}{2}(\alpha+\beta)\ket{{2}}.
\end{align}
The readout process involves an additional photon source, where the emitted photon becomes entangled with the stored spin. At the end of the process, a measurement of the spin reveals the encoded information.\\
\indent Normalizing the state measured in $\ket{+}$ is
\begin{align}
    \ket{\psi_{\rm write}}=\frac{\langle +\vert\psi\rangle}{\mathcal{N}}=\tilde{\alpha}\ket{{1}}+\tilde{\beta}\ket{{2}}
\end{align}
with $\tilde{\alpha}=\frac{1}{2\mathcal{N}}(-\alpha-\beta)$, $\tilde{\beta}=\frac{1}{2\mathcal{N}}(-\alpha+\beta)$ and $\mathcal{N}=\vert\vert\langle +\vert\psi\rangle\vert\vert$. 
We now exemplary perform the reading process step for the state $\ket{+}$. It is
\begin{equation}
\begin{split}
    &\frac{1}{\sqrt{2}}(\ket{e}+\ket{l})(\tilde{\alpha}\ket{{1}}+\tilde{\beta}\ket{{2}})\\
    &\overset{{\rm early\, reflection}}{\rightarrow} \frac{1}{\sqrt{2}}(-\tilde{\alpha}\ket{e{1}}+\tilde{\alpha}\ket{l{1}}+\tilde{\beta}\ket{e{2}}+\tilde{\beta}\ket{l{2}})\\
    &\overset{\pi/2\, {\rm rotation}}{\rightarrow} \frac{1}{\sqrt{2}}(-\frac{\tilde{\alpha}}{\sqrt{2}}(\ket{e{1}}+\ket{e{2}})\\&+\frac{\tilde{\alpha}}{\sqrt{2}}(\ket{l{1}}+\ket{l{2}})+\frac{\tilde{\beta}}{\sqrt{2}}(-\ket{e{1}}\\&+\ket{e{2}})+\frac{\tilde{\beta}}{\sqrt{2}}(-\ket{l{1}}+\ket{l{2}}))\\
    &\overset{{\rm late\, reflection}}{\rightarrow} \frac{1}{\sqrt{2}}(-\frac{\tilde{\alpha}}{\sqrt{2}}(\ket{e{1}}+\ket{e{2}})\\&+\frac{\tilde{\alpha}}{\sqrt{2}}(-\ket{l{1}}+\ket{l{2}})+\frac{\tilde{\beta}}{\sqrt{2}}(-\ket{e{1}}\\&+\ket{e{2}})+\frac{\tilde{\beta}}{\sqrt{2}}(\ket{l{1}}+\ket{l{2}})).
\end{split}
\end{equation}
With the definitions for $\tilde{\alpha}$ and $\tilde{\beta}$ the state reads
\begin{equation}
    \ket{\psi_{\rm read}}=\frac{1}{\mathcal{N}}(\alpha\ket{e}+\beta\ket{l})\ket{{1}}+\frac{1}{\mathcal{N}}(-\alpha\ket{l}+\beta\ket{e})\ket{{2}}.
\end{equation}
Measuring in the $Z$-basis yields the correct photonic qubit.
\section{OPTIMIZATION OF OPTICAL SPIN GATES}\label{app:spingates}
\subsection{Master Equation}\label{app:phonons}
The time evolution of the SnV's spin qubit using a Raman scheme is governed by the Lindblad master equation
\begin{equation}
    \dot{\rho}(t)=-\i[H(t),\rho(t)]+\sum_{k} L_{k}\rho(t) L_{k}^\dagger-\frac{1}{2}\{L_{k}^\dagger L_{k},\rho(t)\}
\end{equation}
\\
where the detailed description of the Hamiltonian $H(t)$ and photonic decay processes using Fermi's golden rule are explained in \cite{pieplow_deterministic_2023}. There is knew knowledge about the detailed modeling of phononic processes \cite{harris_coherence_2023}. According to \cite{harris_coherence_2023} the phononic decay rates for the Lindblad-operators
    \begin{align}
        L_{ij}=\sqrt{\gamma_{ij}}\ket{j}\bra{i}
    \end{align}
    are
    \begin{align}
        \gamma_{ij}=2\pi\sum_R \vert h_{Rij}\vert^2\chi_R\vert \omega_i-\omega_j\vert^3 n(\omega_i-\omega_j),\\R=E_{gx},E_{gy},i,j=1,2,3,4\\
        \gamma_{ij}=2\pi\sum_R \vert h_{Rij}\vert^2\chi_R\vert \omega_i-\omega_j\vert^3 n(\omega_i-\omega_j),\\R=E_{ux},E_{uy},i,j=A,B,C,D
    \end{align}
    with the coefficients
    \begin{align}
        h_{Egx}=-S_g^\dagger(\sigma_x\otimes\mathds{1})S_g,\\ h_{Egy}=-S_g^\dagger(\sigma_y\otimes\mathds{1})S_g,\\
         h_{Eux}=-S_u^\dagger(\sigma_x\otimes\mathds{1})S_u,\\ h_{Euy}=-S_u^\dagger(\sigma_y\otimes\mathds{1})S_u,
    \end{align}
    where $S_b,b=g,u$ refers to the eigenbasis of the SnV in the ground- and excited state, respectively.
    The phononic absorption cross-section is given by
    \begin{align}
        \chi_R=\sum_{m=1}^3\int_S \frac{\hbar{\rm tr}^2(D_R\bm{k}\bm{q}^T_{\bm{k}m})}{16\pi^3\rho c_{\bm{k}m}^5}\,{\rm d}S ~.
    \end{align}
    The integration is performed over the unit sphere $S$, which we perform numerically.
    The phononic occupation number is given by
    \begin{align}
    n(\omega)=\begin{cases}
            (e^{\hbar\vert\omega\vert/k_B T}-1)^{-1},\quad\omega>0\\
            (e^{\hbar\vert\omega\vert/k_B T}-1)^{-1}+1,\quad\omega<0
        \end{cases}
    \end{align}
    where $k_B$ is the Boltzmann constant and  $T$ the temperature.
    
    The strain susceptibility matrices are given by \cite{harris_coherence_2023}
    \begin{align}
        D_{Ebx}=\begin{pmatrix}
            d^b & 0 & \frac{f^b}{2}\\
            0 & -d^b & 0\\
            \frac{f^b}{2} & 0 & 0
        \end{pmatrix},\\ D_{Eby}=\begin{pmatrix}
            0 & -d^b & 0\\
            -d^b & 0 & \frac{f^b}{2}\\
            0 & \frac{f^b}{2} & 0
        \end{pmatrix}
    \end{align}
    with $b=g,u$.
    
    The velocities $c_{\vec q, m}$ are derived from the solution to the eigenvalue problem
    \begin{align}
        \rho\omega_{\bm{k}m}^2 q_i=C_{ijkl}k_j k_k q_l
    \end{align}
    with Hooke's stiffness tensor $\mathbf{C}$ for diamond, the diamond density $\rho=3.51$ ${\rm g}/{\rm cm}^3$, phononic modes $\mathbf{q}$ and wave vector $\mathbf{k}$. The velocities of the phononic modes are
    \begin{align}
    c_{\bm{k}m}=\frac{\omega_{\bm{k}m}}{k}\quad m=1,2,3.
    \end{align}
    Relevant parameters for the computation of the phononic absorption cross-section are listed in Tab. \ref{tab:phonon}.
    \begin{table}[tb]
    \begin{center}
    \caption{Parameters for the phononic decay rates \cite{harris_coherence_2023}}
        \centering
        \begin{tabular}{ccccc}
         \toprule
         $C_{11}$/GPa & $C_{12}$/GPa & $C_{44}$/GPa & $d^g\,(d^u)$/PHz & $f^g\,(f^u)$/PHz\\
         \hline
         $1079.6$ & $126.73$ & $578.16$ & $0.787$ $(0.956)$ & $-0.562$ $(-2.555)$\\
         \toprule
        \end{tabular}
        \label{tab:phonon}
    \end{center}
    \end{table}
\subsection{Optimization}
\begin{table*}[tb]
        \centering        
        \caption{Fidelities for three magnetic field strengths for the case of a pulse train of four $\pi/8$ pulses for achieving a $\pi/2$ gate. The pulse amplitudes are converted to an electric field strength by evaluating $\hbar E_i/ae$ for $i=1,2$ with the quantity $a=55$ pm \cite{pieplow_deterministic_2023} and the elementary charge $e$ at the temperature $T=0.1$ K. A robustness analysis towards disturbances in the optimized parameters is discussed in App. \ref{app:robustness}. The effective gate duration is $T_g=40\sigma$ with $\sigma=\tau_{\pi/8}/(2\sqrt{2\log(2)})$ and the optimized full width half maximum $\tau_{\pi/8}$ of the $\pi/8$ pulse.}
        \begin{tabular}{cccccccccccc}
            \toprule
            $B_{\rm dc}$ (T)  & $4\tau_{\pi/8}$ (ps) & $\Delta_A$ (GHz) & \vspace{.1cm}  $\theta_{\rm dc}$ (deg)  & $\phi_1$ (deg) & $\phi_2$ (deg) & $\theta_1$ (deg) & $\theta_2$ (deg) & $\phi_p$ (deg) & $E_1$ (GHz) & $E_2$ (GHz) & $F_d(\pi/2)$\\
           \hline
           $0.3$ & $151.65$ & $35.73$ & $81.4$ & $84.5$ & $116.35$ & $118.84$ & $92.76$ & $140.69$ & $54.4$ & $56.02$ & $0.9950$\\
           $1.0$ & $64.23$ & $110.53$ & $64.62$ & $97.66$ & $105.31$ & $108.93$ & $103.88$ & $167.31$ & $220.44$ & $223.05$ & $0.9969$\\
           $3.0$ & $353.32$  & $99.66$ & $43.11$ & $100.70$ & $100.63$ & $104.49$ & $104.58$ & $90.03$ & $68.44$ & $72.33$ & $0.9978$\\
           \toprule
        \end{tabular}
        \label{tab:table_ext}
        \end{table*}
We optimize for a $\pi/2$ rotation. The operational fidelity is
\begin{align}
    F_d(\pi/2)={\rm tr}^2 \left(\sqrt{\sqrt{\rho_{\rm tgt}}\rho(T)\sqrt{\rho_{\rm tgt}}}\right)
\end{align}
where $\rho_{\rm tgt}$ is the target state and $\rho(T)$ is the propagated state.
The infidelity is defined as 
\begin{align}
    I_d(\pi/2) = 1 - F_d(\pi/2).
\end{align}
The initial state is $\rho_0=\ket{\psi(t_0)}\bra{\psi(t_0)}$ with
\begin{align}
    \ket{\psi(t_0)} = \frac{1}{\sqrt{2}}(\ket{{1}}\ket{e} + \ket{{2}}\ket{l})~.
    \label{eq:fid_in_state}
\end{align}
The target state is $\rho_{\rm tgt}=R_{\pi/2}\rho_0 R_{\pi/2}^\dagger$
with $R_{\pi/2}=\frac{1}{\sqrt{2}}\begin{pmatrix}
    1 & -1\\ 1 & 1
\end{pmatrix}\otimes\mathds{1}$.\\\\
\indent Throughout the optimizations we fix the magnetic field strength $B_{\rm dc}$ but we optimize its dc field orientation $\theta_{\rm dc}$. Regarding the optical pulse we optimize its full width at half maximum $\tau$, the polarization angles of the pulse $1$ and $2$ which are denoted as $\phi_1,\phi_2,\theta_1,\theta_2$, the phase between the two pulses $\phi_p$ as well as the detuning of the pulses with respect to the lowest lying energy eigenstate in the excited state manifold $\Delta_A$ and their amplitude $E_1$ and $E_2$. Their definition is stated in \cite{pieplow_deterministic_2023}.\\
\indent We perform global optimization in Hilbert space due to numerical costs. In Hilbert space the fidelity is
\begin{align}
    F_f(\theta) &= |\bra{\psi(t_0)}U^\dagger R_{\theta} \ket{\psi(t_0)}|^2
    \label{eq:fidelity}
\end{align}
where the time evolution operator $U$ is calculated by numerically integrating  $\ket{\dot{\psi}} = -{\rm i}H(t)\ket{\psi}$ \cite{Virtanen2020}. 
In \cite{pieplow_deterministic_2023} it is mentioned that the population in higher lying levels must be penalized in order to achieve high fidelity optical spin gates. To achieve this goal, we utilize two key components:
\begin{enumerate}
    \item We introduce a penalty term 
\begin{align}
    P_e=\frac{1}{T}\int_0^T \vert\psi_e (t)\vert^2\,{\rm d}t
\end{align}
with the gate duration $T$ and some excited level $e$ and
\item we apply a pulse train.
\end{enumerate}
We choose a $\pi/8$ rotation around the $y-$axis on the Bloch sphere as the target gate, i.e. 
\begin{align}
    R_{\pi/8}=\cos(\pi/8)\mathds{1}-{\rm i}\sin(\pi/8)\sigma_y.
\end{align}
Empirically, the population in level $e=C$ corresponding to the seventh level inside the eight-level SnV system is most influential which is the reason why we choose that level for the penalty term. The objective function is
\begin{align}
    J=\alpha (1-F_f(\pi/8))+(1-\alpha)P_C, \quad\alpha\in [0,1].
\end{align}
To achieve high fidelity optical spin gates we apply the following procedure:
\begin{enumerate}
    \item Minimize the objective function $J_\alpha$ with the method differential evolution \cite{mayer_differential_2005} for a range of values $\alpha\in [0,1]$.
    \item Evaluate the dissipative infidelity $I_d(\pi/8)$ at the computed minimizer for each $\alpha$ and choose the smallest one. The best found minimizer is called $x_{\rm opt,de}$.
    \item Apply Nelder-Mead's algorithm \cite{olsson_nelder-mead_1975} on the objective function $I_d(\pi/8)$ initializing in $x_0=x_{\rm opt,de}$. The minimizer is called $x_{\rm opt,nm}$.
    \item Evaluate the dissipative $\pi/2$ gate infidelity $I_d(\pi/2)$ in $x_{\rm opt,nm}$ by composing four of the optimized $\pi/8$ gates from the previous steps.
\end{enumerate}
\begin{figure*}[tb]
    \centering
    \includegraphics[width=\linewidth]{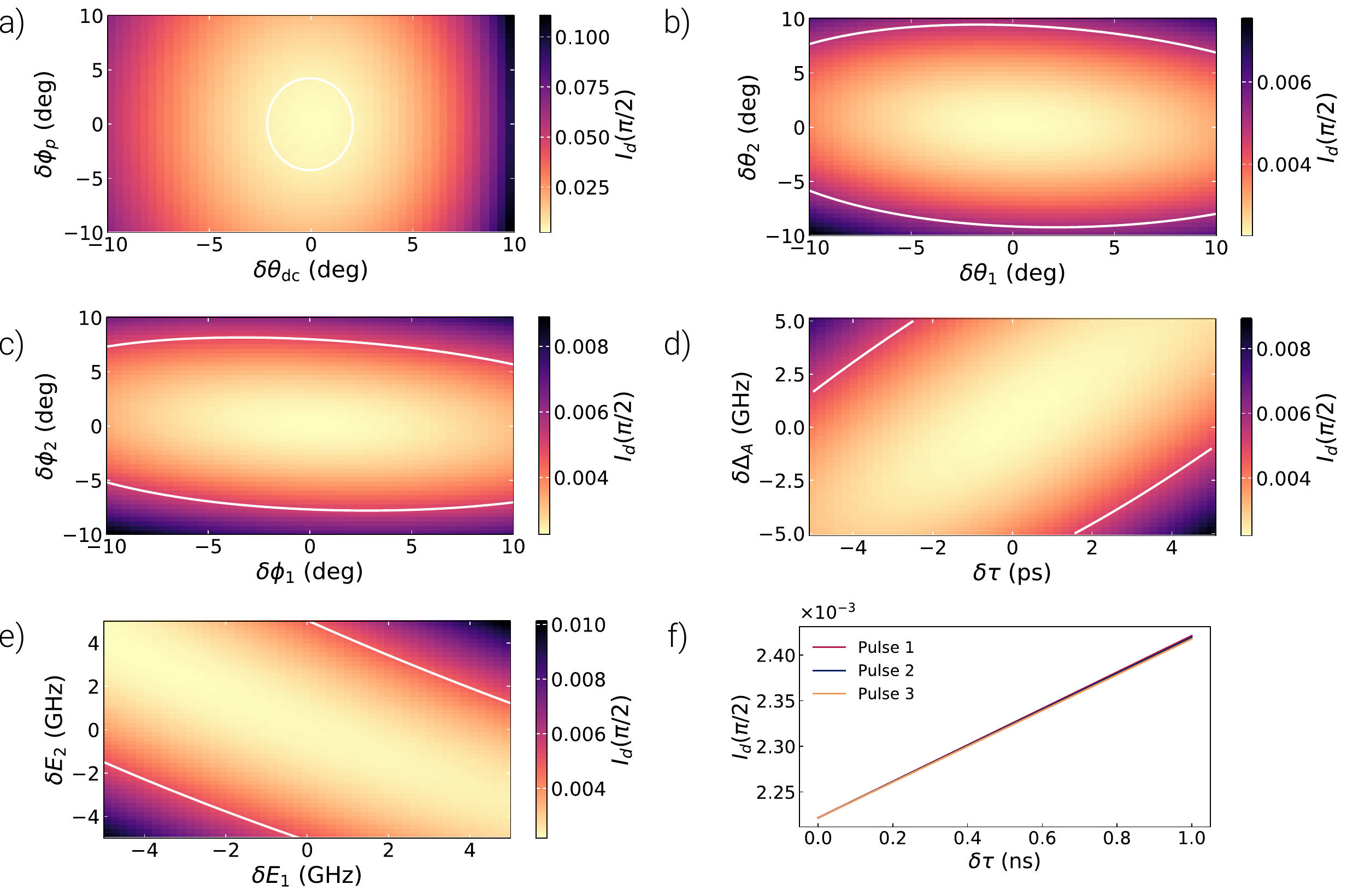}
    \caption{Illustration of the behavior of the infidelity $1-F$ in a local environment around the optimized parameters $(\phi_1,\phi_2)$ (a), $(\phi_p,\theta_{\rm dc})$ (b), $(E_1,E_2)$ (c), $(\Delta_5,\tau)$ (d) $(\theta_1,\theta_2)$ (e) and interpulse delay disturbance $\delta\tau$ (f)  from Tab. \ref{tab:table_ext} for $B_{\rm dc}=3.0$ T. The contour line denotes the infidelity $0.005$.}
    \label{fig:rob}
\end{figure*}
Step 4 is done using linearity of the dynamical map. Consider a propagated basis state subject to a $\pi/8$ gate $\rho_{\pi/8}(T)=\sum_{ij} \rho_{\pi/8,ij} (T)\ket{i}\bra{j}$. To compute the propagated state under a $\pi/2$ gate it is sufficient to apply three additional $\pi/8$ gates, i.e. $\rho_{\pi/2}(T)=\mathcal{D}_{\pi/8}(\mathcal{D}_{\pi/8}(\mathcal{D}_{\pi/8}(\rho_{\pi/8}(T))))$. Using the propagated basis $\mathcal{D}_{\pi/8}(\ket{i}\bra{j})$ for $i,j={1},{2}$ it is straightforward to compute a propagated state under a $\pi/2$ pulse.\\
\indent In Tab. \ref{tab:table_ext} the resulting $\pi/2$ gate fidelities at the temperature $T=0.1$ K and optimized variables at the magnetic field strengths $B_{\rm dc}=0.3$ T, $1.0$ T, $3.0$ T are listed. We find $\pi/2$ gate fidelities exceeding $0.99$ for all three magnetic field strengths which is a promising indicator for a successful use for quantum token applications.\\
\section{ROBUSTNESS OF THE $\pi/2$ GATE}\label{app:robustness}
Since the optimized parameters from Tab. \ref{tab:table_ext} are not exactly experimentally achievable a robustness analysis to uncertainties is provided. We analyze the behavior of the infidelity $I_d(\pi/2)$ in a local environment of the optimized parameters. The optimized point consists of $10$ components, each of which represent a physical quantity. We show five projections: 
\begin{itemize}
    \item Projection 1: $(\theta_{\rm dc},\phi_p)$
    \item Projection 2: $(\theta_1,\theta_2)$
    \item Projection 3: $(\phi_1,\phi_2)$
    \item Projection 4: $(\tau,\Delta_5)$
    \item Projection 5: $(E_1,E_2)$
\end{itemize}
It is important to emphasize that these projections should not be compared against each other because the physical quantities represented by the components shown in Tab. \ref{tab:table_ext} in each projection are fundamentally different. The visualizations attempt to show how small perturbations in specific components influence the objective function. In that way we gain a better understanding about the quantitative behavior of the infidelity in a local environment of the optimized parameters. We also want to identify a region where the infidelity stays below a certain value to provide conclusions about robustness of the $\pi/2$ gate. We define the region of parameters where the infidelity stays below $5\cdot 10^{-3}$ as the robustness region.
In Fig. \ref{fig:rob} that region lies within the contour line. We observe elliptical shapes for the robustness regions for all the projections. For all parameters except the phase $\phi_p$ and magnetic field orientation $\theta_{\rm dc}$ we observe that the robustness region is almost as large as the considered area. In Fig. \ref{fig:rob}a, however, you can see that there is a sensitive dependence on the parameters $\phi_p$ and $\theta_{\rm dc}$ indicating that the error of these parameters must be below approximately $3$ deg for achieving an infidelity below $5\cdot 10^{-3}$.
We also visualize the infidelity as a function of disturbances $\delta\tau$ of the interpulse-delay $T=10\sigma+100$ ps with $\sigma=\tau_{\pi/8}/(2\sqrt{2\log(2)})$ for $B=3.0$ T from Tab. \ref{tab:table_ext}. We choose a maximal disturbance of $\delta\tau=1$ ns because we find such delays in \cite{Rickert2024}. We observe that the infidelity when delaying after pulse 1, 2, 3 stays below $2.45\cdot 10^{-3}$ for $\delta\tau<1$ ns.
The overall conclusion is that the $\pi/2$ gate is robust to uncertainties in the pulse polarizations-, amplitudes-, length and detuning to the lowest lying energy eigenstate and disturbances in the interpulse-delay, however, the phase difference between the pulses and magnetic field orientation must be hit close to the optimized parameters to achieve a sufficiently low infidelity.
\section{OPTIMIZING SPIN-PHOTON CPHASE-GATES}\label{app:cavity}
\begin{figure}
    \centering
    \includegraphics[width=\linewidth]{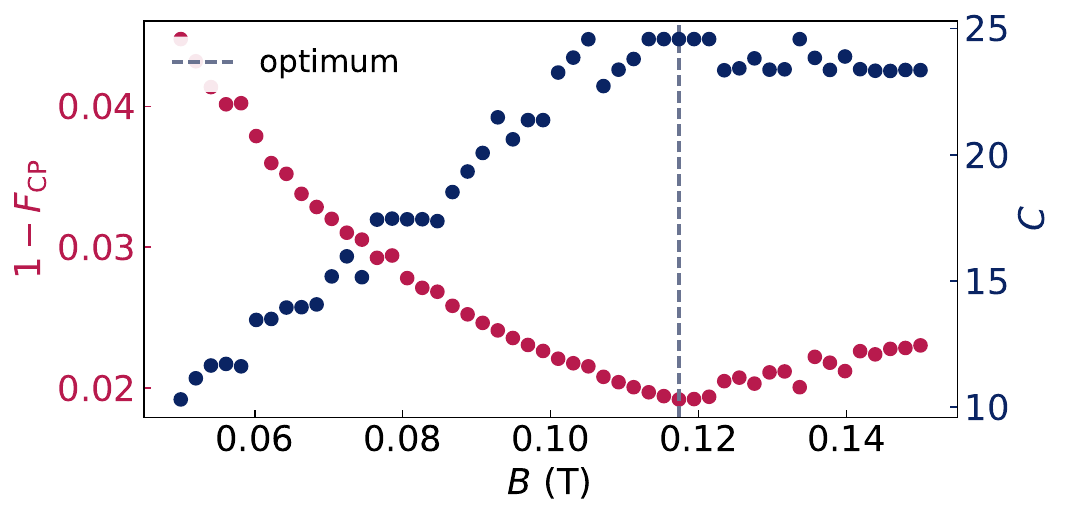}
    \caption{Visualization of the spin-photon entanglement infidelity without crosstalk stated in \cite{strocka_memory_2025} and cooperativity of the transition $\ket{1}\leftrightarrow\ket{A}$ as a function of the magnetic field strength $B_{\rm dc}$ for the orientation $\theta_{\rm dc}=\pi/2$}
    \label{fig:I_C_B}
\end{figure}
\begin{figure}
    \centering
    \includegraphics[width=\linewidth]{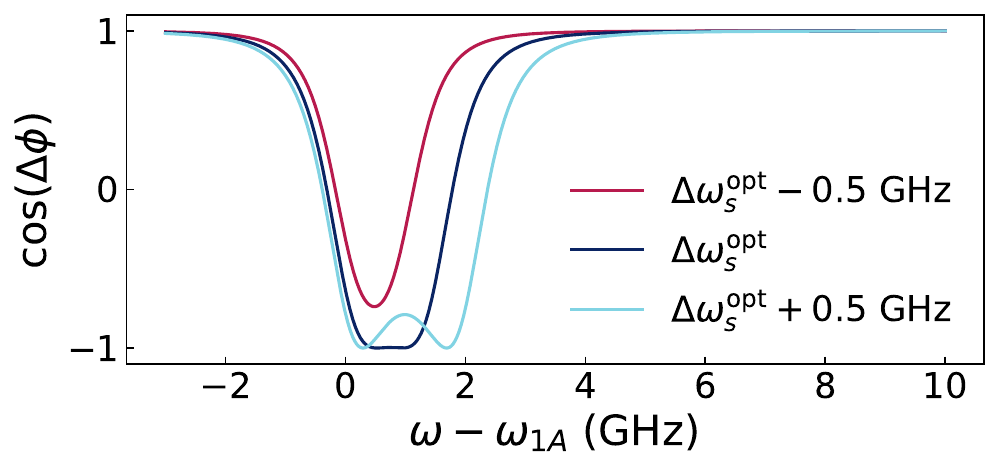}
    \caption{Visualization of the expression $\cos(\Delta\phi)$ with $\Delta\phi:=\phi_1-\phi_2$ and the phase coefficients $\phi_1$ and $\phi_2$ explained in \cite{strocka_memory_2025} describing the phase of the reflected photonic mode when the G4V's spin is initialized in $\ket{1}$ or $\ket{2}$, respectively as a function of the detuning of the incident photon from the atomic transition frequency $\omega-\omega_{1A}$ for the optimal optical splitting $\Delta\omega_s^{\rm opt}$ and two other optical splittings}
    \label{fig:opt_spl}
\end{figure}
For optimizing spin-photon CPHASE-gates necessary to produce spin-photon entanglement we apply two steps:
\begin{enumerate}
    \item Educated guess: Using simplicial homology global optimization \cite{endres_simplicial_2018} we maximize the fidelity stated in \cite{strocka_memory_2025} without crosstalk to compute the triple $(\kappa,\delta_c,\delta_0)$ with $\delta_c=\omega_{1A}-\omega_c$ and $\delta_0=\omega_{1A}-\omega_0$. We apply the Powell method \cite{Powell1964} for local search without crosstalk and further improve the initial guess.
    \item Local search: Using Powell's method \cite{Powell1964} suitable to perform local optimization of the fidelity subject to box-constraints specifically necessary to ensure the cooperativity limit $C=25$ is fulfilled we optimize the fidelity stated in \cite{strocka_memory_2025} subject to the Heisenberg-Langevin equations as derived in \cite{strocka_memory_2025}.
\end{enumerate}
When microwave control is chosen for generating the spin $\pi/2$ rotation there is no limitation on the magnetic field orientation and strength \cite{pieplow_efficient_2024}. The magnetic field orientation and strength are the central ingredients to influence the optical splitting of the SnV \cite{strocka_repeater_2025,omlor_entanglement_2024}. Achieving a spin–photon CPHASE gate for broadband photons benefits from a large optical splitting. This is because increasing the cooperativity flattens the phase spectra \(\phi_1\) and \(\phi_2\) without causing them to overlap, thereby enhancing the gate performance \cite{strocka_repeater_2025,omlor_entanglement_2024}.
However, state-of-the-art G4V-cavities face an inherent upper bound on the cooperativity \cite{Bhaskar2020,Herrmann2024,nina2025}. While this does not lead to a significant fidelity loss for the SiV, it poses a greater challenge for the SnV, particularly for broadband photons \cite{strocka_repeater_2025}. In the present study, however, we focus on narrow-band photons with \(\gamma < 600\ \mathrm{MHz}\), for which cooperativity limitations in group-IV color center cavities are less critical. We adopt a cooperativity bound of \(C = 25\), which has been demonstrated for SiV cavities \cite{Bhaskar2020} but not yet for SnV cavities \cite{Herrmann2024,nina2025}. We consider this value to represent a realistically achievable cooperativity for SnV centers in the near future.

For microwave control we assume $\theta_{\rm dc}=\pi/2$ due to the high efficiency for microwave control \cite{pieplow_efficient_2024}. We further assume that the SnV is embedded in a low-strain environment, which also enhances the optical splitting \cite{strocka_repeater_2025}. Due to the cooperativity limit \(C = 25\), the phase spectra exhibit a finite flatness. This level of flatness is optimal for a specific value of the optical splitting at a given photon bandwidth.
To provide evidence for that statement, we apply the scheme explained in 1. without crosstalk for $\gamma=600$ MHz for magnetic field strengths $B_{\rm dc}\in [0.05,0.15]$ T. In Fig. \ref{fig:I_C_B} we visualize the optimized infidelity $1-F_{\rm CP}$ and cooperativity $C$ on the considered range of magnetic field strengths. We observe a monotonically decreasing infidelity while the cooperativity increases for an increasing magnetic field strength for $B_{\rm dc}<0.12$ T. Within this range of magnetic-field strengths, the optimal cooperativity remains below \(25\), since the limited optical splitting allows the phase spectra to cross. This also explains why the infidelity improves: as the optical splitting increases, enhanced flatness of the phase spectra compensates for the non-zero photon bandwidth \(\gamma\).
Furthermore, we observe that the infidelity begins to increase in a non-monotonic manner for magnetic fields \(B_{\rm dc} > 0.12\ \mathrm{T}\). This behavior arises because the cooperativity limit \(C = 25\) prevents any further improvement once the optical splitting becomes too large. To develop intuition for this effect, we plot in Fig.~\ref{fig:opt_spl} the quantity \(\cos(\Delta\phi)\) with \(\Delta\phi := \phi_1 - \phi_2\), where the phase functions \(\phi_1\) and \(\phi_2\) are defined in \cite{strocka_memory_2025} as the reflection phases of the photonic mode when the G4V spin is initialized in \(\ket{1}\) and \(\ket{2}\), respectively. The curves are shown as a function of the detuning \(\omega - \omega_{1A}\) for the optimal optical splitting \(\Delta\omega_s^{\rm opt}\), as well as for values slightly below and above this optimum.

For \(\Delta\omega_s = \Delta\omega_s^{\rm opt} - 0.5\ \mathrm{GHz}\), the condition \(\cos(\Delta\phi(\omega)) = -1\) is not satisfied for any frequency. At the optimal optical splitting \(\Delta\omega_s^{\rm opt}\), corresponding to a magnetic field \(B \approx 0.12\ \mathrm{T}\) at \(\theta_{\rm dc} = \pi/2\), the phase condition is fulfilled over a broad frequency interval. That is because the phase jumps of $\phi_1$ and $\phi_2$ overlap. Both spins produce the desired phase change in almost the same frequency region. The optical splitting is increased further, the broad flat region breaks into two isolated and narrow frequency points at which the phase condition can still be met. That happens because the phase jumps no longer overlap.
This reduction in width of the frequency points explains the rise in infidelity observed in Fig.~\ref{fig:I_C_B} for \(B > 0.12\ \mathrm{T}\) and provides the intuition of an optimal optical splitting at a given incident photon bandwidth.

We now optimize subject to crosstalk as stated in \cite{strocka_repeater_2025} but assuming $C=25$ and that the magnetic field strength is an optimization variable. We apply Powell's method \cite{Powell1964} initialized in the educated guess $(B_{\rm dc},\delta_c,\delta_0)=(0.121\,{\rm T},-372\,{\rm MHz}, -839\,{\rm MHz})$ with an infidelity of $1-F_{\rm CP}=0.0192$ without crosstalk and $0.044$ with crosstalk. We find the result stated in Tab. \ref{tab:CP_data}. We observe that Powell's method \cite{Powell1964} improved the infidelity to $1-F_{\rm CP}=0.0204$ and found the magnetic field strength $B_{\rm dc}=0.139$ T.

When implementing the Raman scheme to realize a spin \(\pi/2\) rotation, the magnetic-field strength and orientation are constrained to the values listed in Tab.~\ref{tab:table_ext}.
We apply the steps 1. and 2. for the magnetic field strengths and orientations stated in Tab. \ref{tab:table_ext} for the incident's photon bandwidth $\gamma=600$ MHz. The resulting triple $(\kappa,\delta_c,\delta_0)$ as well as infidelity and efficiency are listed in Tab. \ref{tab:CP_data}. We observe that the infidelities are higher than those at the optimal magnetic-field configuration discussed above, which is consistent with the reasoning provided earlier.

\begin{figure}
    \centering
    \includegraphics[width=\linewidth]{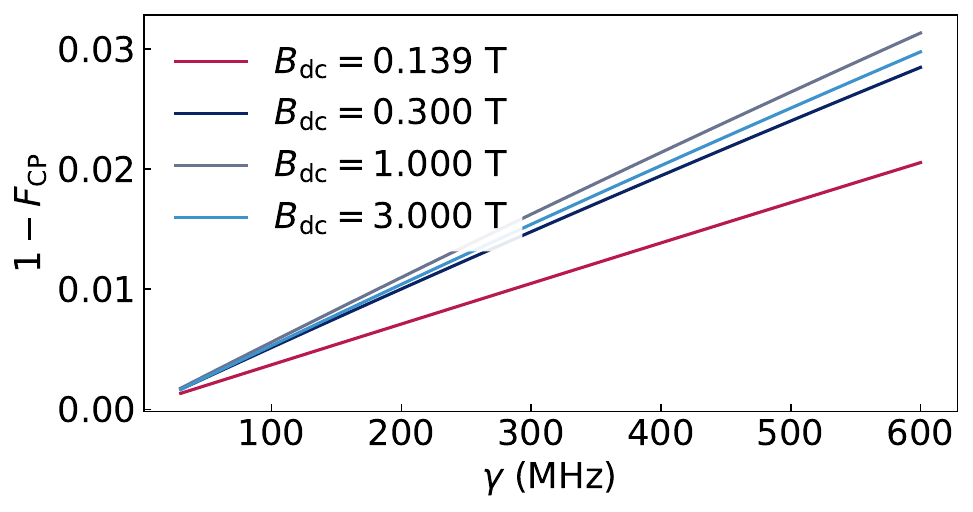}
    \caption{Illustration of the infidelity as a function of the incident's photon bandwidth for the optimized parameters listed in Tab. \ref{tab:CP_data}. The efficiency is invariant on the considered bandwidth range and also listed in Tab. \ref{tab:CP_data}.}
    \label{fig:I_ga}
\end{figure}

In Fig. \ref{fig:I_ga} we visualize the infidelity as a function of the bandwidth for the configurations shown in Tab. \ref{tab:CP_data}. We observe that the infidelity improves as the photon gets narrower. That is intuitive following the explanation from \cite{omlor_entanglement_2024}. Crosstalk does not change that behavior.

\begin{figure*}
    \centering
    \includegraphics[width=\linewidth]{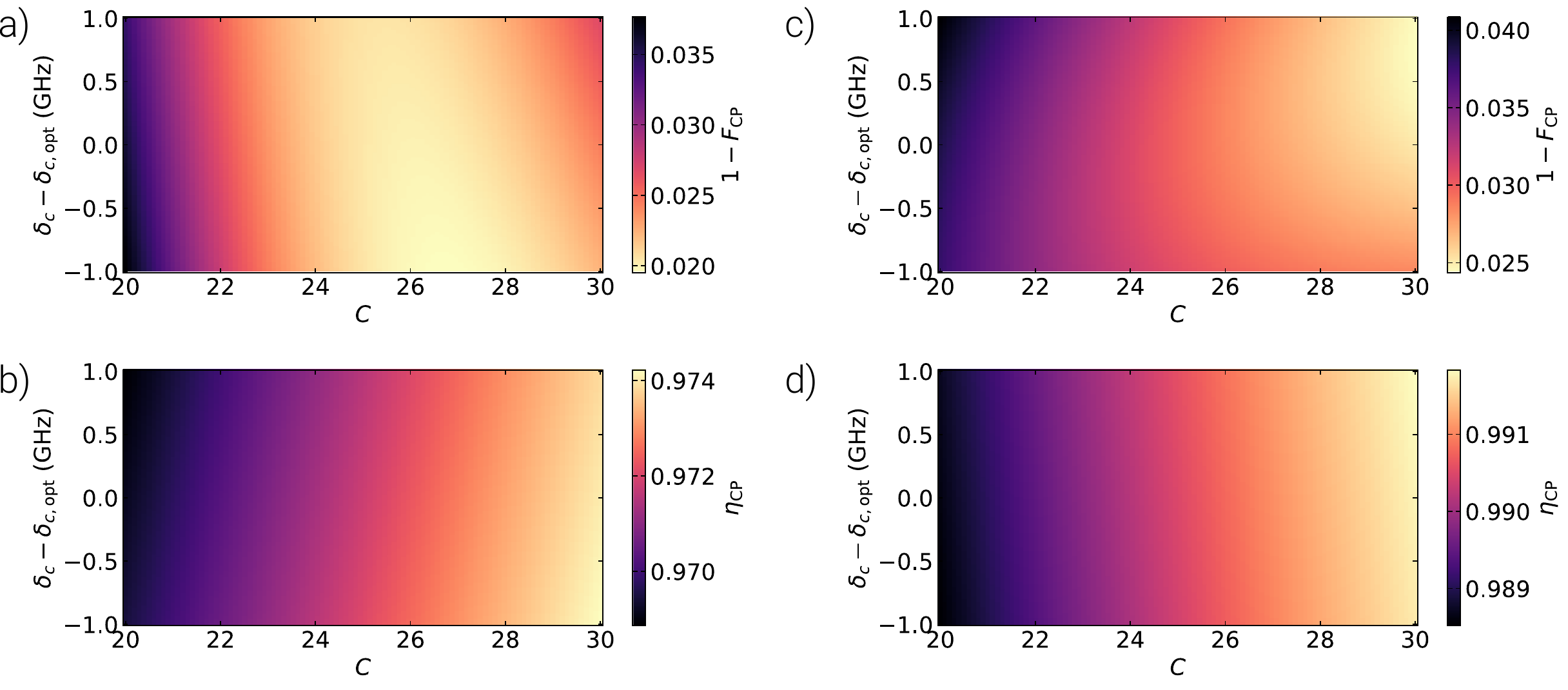}
    \caption{Infidelity $1-F$ and efficiency $\eta$ as a function of the cooperativity $C$ between the transitions $\ket{1}$ and $\ket{A}$ and variations in the detuning of the cavity mode central frequency $\delta_c$ for $B_{\rm dc}=0.139$ T in a), b) and $B_{\rm dc}=3.0$ T in c), d) with the corresponding parameters listed in Tab. \ref{tab:CP_data}}
    \label{fig:sens}
\end{figure*}

In Fig. \ref{fig:sens} we visualize the infidelity and efficiency as a function of variations in the cavity mode central frequency detuning and cooperativity for $B_{\rm dc}=0.139$ T and $B_{\rm dc}=3.0$ T with the parameters from Tab. \ref{tab:CP_data}. We observe that both the infidelity and efficiency vary only slightly within the examined parameter ranges. Furthermore, higher cooperativities reduce the infidelity for \(B_{\rm dc} = 3.0\ \mathrm{T}\), which is intuitive since the increased cooperativity flattens the phase spectrum without causing overlap. In contrast, for \(B_{\rm dc} = 0.139\ \mathrm{T}\), the optical splitting is so small that further flattening of the spectra leads to overlap, resulting in a drop in fidelity. Overall, we conclude that the infidelity and efficiency are robust against perturbations in the cavity-mode central frequency and cooperativity.

\begin{table*}[tb]
        \centering        
        \caption{Optimized triple $(\kappa,\delta_c,\delta_0)$ with respective infidelity $1-F_{\rm CP}$ and efficiency $\eta_{\rm CP}$ for the magnetic field strength $B=0.139,0.3,1.0,3.0$ T at the bandwidth $\gamma=600$ MHz. The cooperativity for the transition $\ket{1}\leftrightarrow\ket{A}$ is $C=25$ for the listed cases.}
        \begin{tabular}{ccccccc}
            \toprule
            $B_{\rm dc}$ (T) & $\theta_{\rm dc}$ (deg) & $\kappa$ (GHz) & $\delta_c$ (GHz) & $\delta_0$ (GHz) & $\eta_{\rm CP}$ & $1-F_{\rm CP}$\\
           \hline
           $0.139$ & $90$ & $42.14$ & $-7.85$ & $-0.31$ & $0.9718$ & $0.0204$\\
           $0.3$ & $81.4$ & $42.12$ & $0.08$ & $-3.00$ & $0.9874$ & $0.0285$\\
           $1.0$ & $64.62$ & $42.06$ & $-5.40$ & $-5.63$ & $0.9892$ & $0.0313$\\
           $3.0$ & $43.11$ & $42.14$  & $-5.38$ & $-5.66$ & $0.9895$ & $0.0298$\\
           \toprule
        \end{tabular}
        \label{tab:CP_data}
    \end{table*}
\section{PERFORMANCE OF STORAGE AND RETRIEVAL}\label{app:performance}
To evaluate the performance of storage and retrieval we evaluate the average efficiency and infidelity for the storage and retrieval process of a photonic qubit from the basis states necessary for Wiesner's scheme $\mathcal{B}=\{\ket{e},\ket{l},\ket{+},\ket{-}\}$ with $\ket{+}=(\ket{e}+\ket{l})/\sqrt{2}$ and $\ket{-}=(\ket{e}-\ket{l})\sqrt{2}$ \cite{wiesner_conjugate_1983}. Following the Kraus-formalism shown in \cite{strocka_memory_2025} we evaluate the stored spin states $\rho_+(0)$ and $\rho_-(0)$ for all $\ket{\psi_x}\in\mathcal{B}$. Depending on the storage time $T_s$ we propagate the stored spin states defined by $\rho_+(T_s),\rho_-(T_s)$. For both states we perform the retrieval step leading to two measurement outcomes for both states, i.e. $\rho_{+,1},\rho_{+,2},\rho_{-,1},\rho_{-,2}$. Based on these results we construct the complete state as follows
\begin{align}
    \rho_{x,+}=\rho_{+,1}+\sigma_z\sigma_x\rho_{+,2}\sigma_x\sigma_z,\\
    \rho_{x,-}=\rho_{-,1}+\sigma_z\sigma_x\rho_{-,2}\sigma_x\sigma_z
\end{align}
with $\sigma_x=\begin{pmatrix}
    0 & 1 \\
    1 & 0
\end{pmatrix}$ and $\sigma_z=\begin{pmatrix}
    1 & 0 \\
    0 & -1
\end{pmatrix}$.
From these states the total photonic state after read-out is given by
\begin{align}\label{eq:tot_state}
    \tilde{\rho}_x=\rho_{x,+}+\sigma_z\rho_{x,-}\sigma_z.
\end{align}
The efficiency of the joint read-in and read-out process is subsequently given by $\eta_x={\rm tr}(\tilde{\rho}_x)$ and the fidelity is $F_x=\frac{1}{\eta_x}\langle\psi_x\vert\tilde{\rho}_x\vert\psi_x\rangle$ for $x=\pm,e,l$.

\section{SPECTRAL DIFFUSION OF THE EMITTER}\label{app:diff}
In a realistic setting the photon source has some spectral diffusion due to environmental charge noise. We assume a gaussian distribution of the photon's central frequency. The distribution is
\begin{align}
   G(\Delta\nu)=\frac{1}{\sqrt{2\pi}\sigma}e^{-\frac{\Delta\nu^2}{2\sigma^2}}
\end{align}
with a standard deviation $\sigma$ and difference between the ideal central frequency and the shifted one due to spectral diffusion $\Delta\nu$. To quantify the impact of spectral diffusion on the token acceptance rate $\gamma_a$ stated in Eq. \eqref{eq:acceptance_rate} we visualize its dependence on the standard deviation $\sigma$ of the gaussian distribution and the incident's photon bandwidth $\gamma$ for token storage and retrieval. The evaluation of the token acceptance rate $\gamma_a$ requires the average fidelity $\langle F\rangle=1/4 (F_++F_-+F_e+F_l)$ and efficiency $\langle\eta\rangle=1/4 (\eta_++\eta_-+\eta_e+\eta_l)$ where $F_x, \eta_x$ for $x=\pm,e,l$ accounts for both the token storage and retrieval processes. These quantities are given by
\begin{align}
    F_x=1/\eta_x \langle\psi_x\vert\langle\rho_x\rangle\vert\psi_x\rangle,\quad \eta_x={\rm tr}(\langle\rho_x\rangle)
\end{align}
with
\begin{align}\label{eq:state_avg}
    \langle\rho_x\rangle=\int_\mathbb{R} G(\omega-\omega_0)\rho_x(\omega)\,{\rm d}\omega
\end{align}
and the state $\rho_x(\omega)$ the photonic qubit state after storage and retrieval defined in Eq. \eqref{eq:tot_state} assuming the incident's photon frequency $\omega$. To evaluate the state shown in Eq. \eqref{eq:state_avg} for a fixed bandwidth $\gamma$ we compute the state $\rho_x(\omega)$ on an equidistant grid of incident photon's frequencies on the range $\omega\in [\omega_0-\Delta,\omega_0+\Delta]$ with $\Delta=1$ GHz based on the imperfect spin $\pi/2$ roation produced with optical control and the full nonlinear set of Heisenberg-Langevin equations \cite{strocka_memory_2025} modeling spin-photon interaction including crosstalk between undesired transitions. Subsequently, we interpolate each entry of $\rho_x(\omega)$ for $x=\pm,e,l$ with cubic splines \cite{Virtanen2020}. Using numerical integration from \cite{Virtanen2020} we are afterwards able to evaluate the integral shown in Eq. \eqref{eq:state_avg}. Based on the observation that the infidelity is a linear function of the bandwidth of the incident photon which can be deduced from Fig. \ref{fig:I_ga} we evaluate the the state $\rho_x(\omega)$ for the bandwidth $\gamma_1=30$ MHz and $\gamma_2=600$ MHz. Subsequently, we apply a linear interpolation of the infidelity for each $\omega\in [\omega_0-\Delta,\omega_0+\Delta]$. For the efficiency we do the same procedure. However, it does not change as a function of the bandwidth.

\begin{figure}
    \centering
    \includegraphics[width=\linewidth]{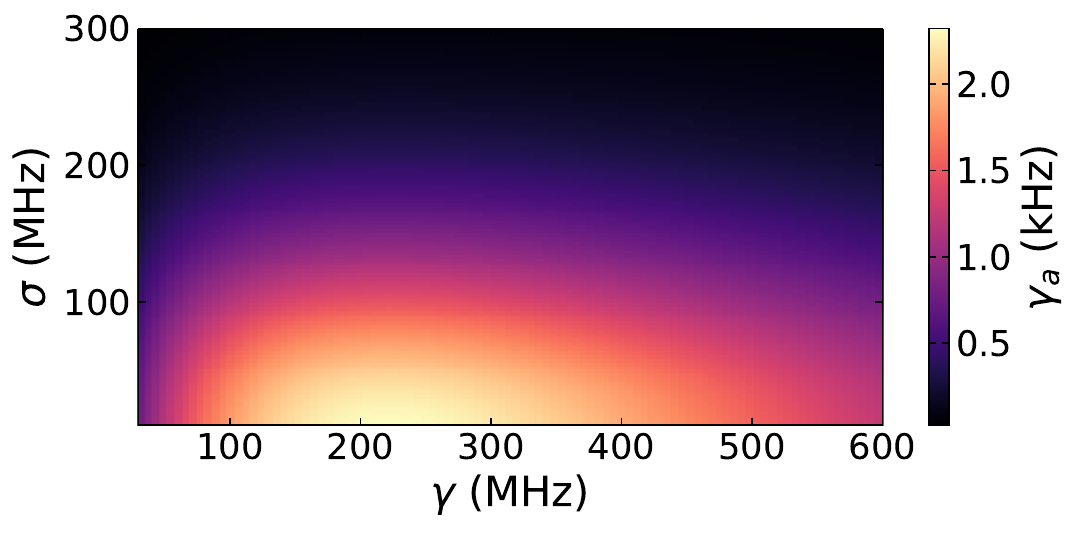}
    \caption{Spectral diffusion's standard deviation $\sigma$ impact on $\gamma_a$ for different  $\gamma$ of the incoming photons is quantified. We assume $L=0.5$ km, $L_{\rm att}=20$ km, $T=0.1$ K, $p_{\rm th}=10^{-4}$, $\eta_c=1$, a pulse train of optical $\pi/8$ pulses to achieve a $\pi/2$ gate for $B_{\rm dc}=3.0$ T (see details in Tab. \ref{tab:table_ext} from App. \ref{app:spingates}), the electron spin as the quantum memory, $T_s=0$ and a gaussian shape of the spectral diffusion with standard deviation $\sigma$. }
    \label{fig:diff_opt}
\end{figure}

Based on the described procedure we evaluate the token acceptance rate $\gamma_a$ as a function of the incident's photon bandwidth $\gamma$ and standard deviation of the gaussian distribution $\sigma$ assuming optical spin control to achieve a $\pi/2$ rotation. We visualize the result in Fig. \ref{fig:diff_opt}. We observe that the acceptance rate $\gamma_a$ only weakly changes for $\sigma<100$ MHz. We conclude that emitters fulfilling lying in that range like for example the SnV itself \cite{trusheim_transform-limited_2020} are appropriate emitters for that scheme.
\\
\section{DECOHERENCE OF THE SAVED STATE}\label{app:spin_dec}
\subsection{Electronic Spin}
When the state is saved in the quantum memory there is dephasing. We model this with a two-level system governed by the Lindblad master equation in the rotating frame
\begin{align}
    \dot{\rho}&=\sum_{k=1}^2 L_k\rho L_k^\dagger-\frac{1}{2}\{L_k^\dagger L_k,\rho\},\\
    \rho(0)&=\rho_0
\end{align}
with the Lindblad operators
\begin{align}
    L_1=\sqrt{\gamma_-}\sigma_-,\quad L_2=\sqrt{\gamma_+}\sigma_+
\end{align}
where $\sigma_-$ and $\sigma_+$ describe the rising and lowering operator respectively and the initial state
\begin{align}
    \rho_0=\begin{pmatrix}
        \rho_{00} & \rho_{01}\\
        \rho_{10} & \rho_{11}
    \end{pmatrix}
\end{align}
which is the state saved in the color center's spin qubit derived in App. \ref{app:performance}. Solving the master equation with that initial state yields
\begin{align}
    \rho(t)&=\begin{pmatrix}
        \rho_{00}(t) & \rho_{01}(t)\\ \rho_{10}(t) & \rho_{11}(t)
    \end{pmatrix},\\
    \rho_{00}(t)&=\left(\rho_{00}-\frac{\gamma_-}{\gamma_- +\gamma_+}\right)e^{-\frac{\gamma_- +\gamma_+}{2}t}+\frac{\gamma_-}{\gamma_- +\gamma_+},\\
    \rho_{01}(t)&=e^{-\frac{\gamma_-+\gamma_+}{2}t}\rho_{01},\\
    \rho_{10}(t)&=e^{-\frac{\gamma_-+\gamma_+}{2}t}\rho_{10},\\
    \rho_{11}(t)&=1-\rho_{00}(t).
\end{align}
\subsection{Nuclear Spin}
If the nuclear spin is considered to be the memory we model pure dephasing using the lindblad-operator $L=\sqrt{\gamma_d}\sigma_z$ with the dephasing rate $\gamma_d=1/T_d$ and $T_d=1$ s \cite{Grimm2025}. The propagated spin state reads
\begin{align}
    \rho(t)=\begin{pmatrix}
        \rho_{00} & \rho_{01}e^{-2\gamma_d t}\\
        \rho_{10}e^{-2\gamma_d t} & \rho_{11}
    \end{pmatrix}.
\end{align}
\end{appendix}
\end{document}